\documentclass[12pt]{article}
\usepackage{epsf}

\newcommand{\bmat}{\left(\begin{array}}
\newcommand{\emat}{\end{array}\right)}

\def\yzero{\smash{\hbox{$y\kern-4pt\raise1pt\hbox{${}^\circ$}$}}}

\def\a{\alpha}

\def\d{\delta}
\def\beq{\begin{equation}}
\def\eeq{\end{equation}}
\def\beqa{\begin{eqnarray}}
\def\eeqa{\end{eqnarray}}

\def\th{\theta}
\def\vt{\vartheta}

\def\-{\hphantom{-}}
\def\ov{\overline}
\def\s2{\frac{1}{\sqrt2}}

\def\beq{\begin{equation}}
\def\eeq{\end{equation}}
\def\beqa{\begin{eqnarray}}
\def\eeqa{\end{eqnarray}}
\def\tr{{\rm tr \,}}

\def\IF{\relax{\rm I\kern-.18em F}}
\def\II{\relax{\rm I\kern-.18em I}}
\def\IP{\relax{\rm I\kern-.18em P}}
\def\IC{\relax\hbox{\kern.25em$\inbar\kern-.3em{\rm C}$}}
\def\IR{\relax{\rm I\kern-.18em R}}

\def\Dsl{\,\raise.15ex\hbox{/}\mkern-13.5mu D} 
\def\IZ{Z\kern-.4em  Z}

\def\k{\kappa}
\def\vt{\vartheta}

\def\eps{\epsilon}

\def\im{{\rm Im}\,}

\newcommand{\drawsquare}[2]{\hbox{%
\rule{#2pt}{#1pt}\hskip-#2pt
\rule{#1pt}{#2pt}\hskip-#1pt
\rule[#1pt]{#1pt}{#2pt}}\rule[#1pt]{#2pt}{#2pt}\hskip-#2pt
\rule{#2pt}{#1pt}}

\newcommand{\fund}{\raisebox{-.5pt}{\drawsquare{6.5}{0.4}}}
\newcommand{\antifund}{\overline{\fund}}

\catcode`\@=11   
\newdimen\@rotdimen
\newbox\@rotbox  

\def\@vspec#1{\special{ps:#1}}
\def\@rotstart#1{\@vspec{gsave currentpoint currentpoint translate
   #1 neg exch neg exch translate}}
\def\@rotfinish{\@vspec{currentpoint grestore moveto}}
\def\@rotr#1{\@rotdimen=\ht#1\advance\@rotdimen by\dp#1%
   \hbox to\@rotdimen{\hskip\ht#1\vbox to\wd#1{\@rotstart{90 rotate}%
   \box#1\vss}\hss}\@rotfinish}
\def\@rotl#1{\@rotdimen=\ht#1\advance\@rotdimen by\dp#1%
   \hbox to\@rotdimen{\vbox to\wd#1{\vskip\wd#1\@rotstart{270 rotate}%
   \box#1\vss}\hss}\@rotfinish}%
\def\@rotu#1{\@rotdimen=\ht#1\advance\@rotdimen by\dp#1%
   \hbox to\wd#1{\hskip\wd#1\vbox to\@rotdimen{\vskip\@rotdimen
   \@rotstart{-1 dup scale}\box#1\vss}\hss}\@rotfinish}%
\def\@rotf#1{\hbox to\wd#1{\hskip\wd#1\@rotstart{-1 1 scale}%
   \box#1\hss}\@rotfinish}%
\def\rotate{\@ifnextchar[{\@rotate}{\@rotate[l]}}
\def\@rotate[#1]#2{\setbox\@rotbox=\hbox{#2}\@nameuse{@rot#1}\@rotbox}

\catcode`\@=12

\topmargin
-1.5cm
\textwidth
15.5cm
\textheight
23.5cm
\oddsidemargin
0.7cm
\evensidemargin
1.2cm

\begin{document}

\makeatletter
\@addtoreset{equation}{section}
\makeatother
\renewcommand{\theequation}{\thesection.\arabic{equation}}
\pagestyle{empty}
\rightline{ IFT-UAM/CSIC-06-45}
\rightline{ CERN-PH-TH/2006-199}
\rightline{\tt hep-th/0609213}
\vspace{1.5cm}
\begin{center}
\LARGE{ Neutrino Majorana Masses\\ from  String Theory Instanton Effects   
\\[10mm]}
\large{ L.E. Ib\'a\~nez$^{1}$ and A. M. Uranga$^{2}$ \\[6mm]}
\small{
${}^{1}$ Departamento de F\'{\i}sica Te\'orica C-XI
and Instituto de F\'{\i}sica Te\'orica  C-XVI,\\[-0.3em]
Universidad Aut\'onoma de Madrid,
Cantoblanco, 28049 Madrid, Spain \\[2mm] and 
\\[2mm]
${}^{2}$ PH-TH Division, CERN \\
CH-1211 Geneva 23, Switzerland\\
(On leave from IFT, Madrid, Spain)\\
[7mm]}
\small{\bf Abstract} \\[12mm]
\end{center}
\begin{center}
\begin{minipage}[h]{16.0cm}
Finding a plausible origin for right-handed neutrino 
Majorana masses in semirealistic compactifications of string theory
remains one of the most difficult problems in string phenomenology.
We argue that right-handed neutrino Majorana masses  are induced by 
non-perturbative instanton effects in certain classes of string 
compactifications in which the $U(1)_{B-L}$ gauge boson has a St\"uckelberg 
mass. The induced operators are of the form $e^{-U}\nu_R\nu_R$ where $U$ is 
a closed string modulus whose imaginary part transforms appropriately 
under $B-L$. This mass term may be quite large since this is not a gauge 
instanton and $Re\,U$ is not directly related to SM gauge couplings.
Thus the size of the induced right-handed neutrino masses could be a few 
orders of magnitude below the string scale, as phenomenologically 
required. It is also argued that this origin for neutrino masses
would predict the existence of R-parity in SUSY versions of the SM. 
Finally we comment on other phenomenological applications of similar 
instanton effects, like the generation of a $\mu$-term, or of Yukawa 
couplings forbidden in perturbation theory.
\end{minipage}
\end{center}
\newpage
\setcounter{page}{1}
\pagestyle{plain}
\renewcommand{\thefootnote}{\arabic{footnote}}
\setcounter{footnote}{0}


\section{Introduction}

In recent years our experimental knowledge about neutrino masses has 
substantially improved. The evidence from solar, atmospheric, reactor and 
accelerator experiments indicates that neutrinos are massive. The observed 
structure of masses and (large) mixings of neutrinos is quite peculiar 
and different from their charged counterparts. The simplest explanation 
for the smallness of neutrino masses is the celebrated see-saw mechanism \cite{seesaw}. 
If there are right-handed neutrinos $\nu_R^a$ with large Majorana masses 
$M_M$ and standard Dirac masses $M_D$, the lightest eigenvalues have 
masses of order
\beq
m_{\nu}\  \simeq \ \frac {M_D^2}{M_M}  \ .
\eeq
which are of order the experimental results for $M_D$ of order of 
standard charged leptons and $M_M\propto 10^{10}-10^{13}$ GeV.

Dirac masses are  expected to be given by standard Yukawa couplings so 
from this point of view they are  tied down to the usual flavor problem of 
the SM, the not yet understood structure of fermion masses of mixings. On 
the other hand the origin of the large Majorana masses for right-handed 
neutrinos is even more mysterious. A natural setting for such masses 
seems to be left-right extensions of the SM like $SO(10)$ unification. 
However in this case the appropriate Higgs fields leading to Majorana 
masses have dimension 126, making the models unattractive. Alternatively 
one may resort to non-renormalizable couplings to 16-plets of Higgs 
fields, but in SUSY models this generically breaks R-parity spontaneously, 
giving  rise to Baryon- and Lepton-number violation (and hence fast 
proton decay) unless one invokes extra protecting symmetries. 

The situation for Majorana masses in the case of string theory is worst 
(for a recent discussion see e.g. \cite{Giedt:2005vx} and references 
therein)
because there is less freedom to play around with models. One of the 
reasons is that Higgs fields with the appropriate quantum numbers to 
couple to the $\nu_R \nu_R$ bilinears at the renormalizable level do not 
appear in any of the semirealistic models constructed up to now. Although 
such couplings may appear at the non-renormalizable level, it is still 
typically problematic to obtain them without at the same time inducing (at 
least in the SUSY case) dangerous B/L-violating couplings. We think it is 
fair to say that there is at present no semirealistic model in which a 
large Majorana mass for the right-handed neutrinos appears in a natural way.

In this paper we present an elegant mechanism for the generation of 
right-handed neutrino masses in string theory. We claim that, in a 
(presumably large but) restricted class of string compactifications with 
semirealistic SM or MSSM light spectrum, there exist string theory 
instanton effects which induce a Majorana mass term for the
right-handed neutrino. They are of the form 
\beq
 \ e^{-\frac {1}{g^2(U)}}\ M_{String}\ c_{ab} \nu_R^a\nu_R^b
\label{maj}
\eeq
%
\begin{figure}
\epsfysize=6cm
\begin{center}
\leavevmode
\epsffile{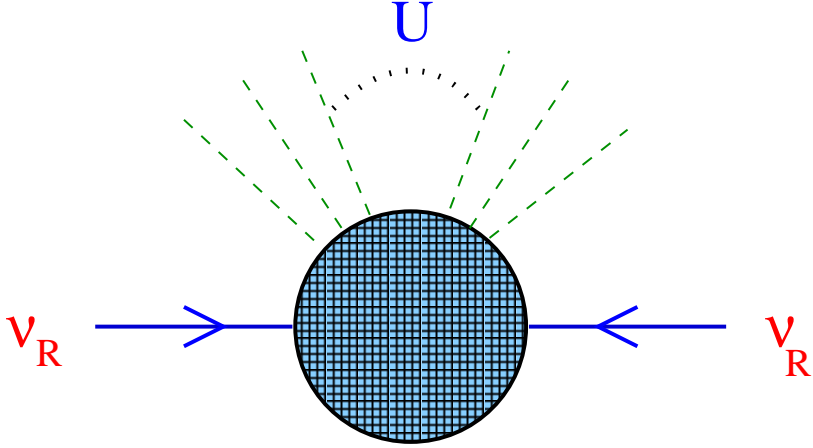}
\end{center}
\caption{Instanton induced right-handed neutrino Majorana mass term.}
\label{instanton4}
\end{figure}
%
The instanton effect is depicted in Figure \ref{instanton4}. Here $U$ 
denotes a set of string moduli, on which the strength $g^2(U)$ of the  
non-perturbative effect depends. Here $g^2$ is {\it not} directly related 
to the SM gauge couplings so that these masses may be quite big although 
naturally suppressed with respect to the string scale $M_{String}$.
A crucial ingredient for the mechanism to work is that the model should 
contain a gauged $B-L$ $U(1)$ symmetry beyond the SM gauge group, whose 
gauge boson gets a St\"uckelberg mass by combining with some scalar modulus 
field (e.g., a RR-scalar in Type II compactifications). Then under some  
conditions  to be discussed below, certain (non-gauge) instanton effects 
analogous to those discussed in \cite{Ganor:1996pe,mcgreevy} give rise 
to a  right-handed Majorana mass term of the above form.

As we said, it is important that $U(1)_{B-L}$ gets a St\"uckelberg mass.
It is a familiar fact in string models that $U(1)$ gauge fields with 
triangle anomalies canceled by the Green-Schwarz mechanism get 
St\"uckelberg masses. On the other hand, as emphasized in \cite{imr} (see 
also \cite{Antoniadis:2002cs}) anomaly-free $U(1)$'s like  $U(1)_{B-L}$ 
may also get St\"uckelberg masses, also by $B\wedge F$ couplings to 
suitable 2-form fields. For instance, a class of SM-like compactifications 
in which $U(1)_{B-L}$ gets a St\"uckelberg mass was provided in \cite{imr}, 
based on models of intersecting D6-branes on an orientifold of type IIA 
on $T^2\times T^2\times T^2$ \footnote{These models are 
non-supersymmetric but a number of $N=1$ SUSY models with MSSM-like 
spectrum in which $U(1)_{B-L}$ gets a St\"uckelberg mass were reported
in \cite{schell}.}. In this paper we use analogous examples to 
illustrate explicitly that (non-gauge) euclidean D2-brane instantons can 
give rise to Majorana mass terms as above. 

On the other hand the mechanism is quite general and works in complete 
analogy in other string compactifications with D-branes (including 
non-geometric CFT compactifications like the models in 
\cite{schell}), or even in heterotic compactifications 
with $U(1)$ bundles. A difference in the heterotic case is that 
the effects can originate from world-sheet instantons, and are hence tree 
level in $g_s$ (and non-perturbative in $\alpha'$).

\medskip

This paper is organized as follows. In Section \ref{scheme} we lay down 
the general idea of generating Majorana mass terms via non-perturbative 
instanton effects in string theory. In section \ref{remarks} we motivate 
the proposal, and in section \ref{instantonmech} we describe the instanton 
induced operator, its symmetry properties, and the microscopic mechanism 
in which it is generated. In section \ref{extra} we discuss some 
additional general aspects of the mechanism. Section \ref{example} 
provides an explicit example of a semirealistic string theory D-brane 
model, where Majorana mass terms arise from D2-brane instantons. Section 
\ref{muterm} discusses the use of similar non-perturbative instanton 
effects in generating other interesting operators, in particular the 
$\mu$-term in supersymmetric models, or Yukawa couplings forbidden in 
perturbation theory. Section \ref{discussion} contains our final comments. 

Appendices contain technical details and additional examples. Appendix 
\ref{generalinst} describes instanton induced operators for completely 
general D-branes models (including type IIB with magnetized branes, 
D-branes at singularities, or even non-geometric CFT compactifications), 
and for heterotic models. Appendix \ref{moremodels} contains an additional
class of semirealistic models allowing for instanton induced Majorana mass 
terms, while appendix \ref{muexample} contains  examples of a 
semirealistic SUSY model allowing for an instanton induced $\mu$-term.

\medskip

As this paper was ready for submission, ref. \cite{Blumenhagen:2006xt}
appeared, which also discusses non-perturbative instanton effects in 
semirealistic string models.

\section{The general scheme}
\label{scheme}

\subsection{General remarks}
\label{remarks}

The discussion of the physics of neutrino masses in string models should 
clearly be carried out within the setup of semi-realistic string 
constructions reproducing structures close to the (possibly 
supersymmetric) Standard Model. It is interesting to point out that the 
presence of right-handed neutrinos is a quite generic feature within this 
class. For instance, in type II compactifications with D-branes, 
right-handed neutrinos arise from open strings stretched between  two 
stacks of $U(1)$ branes. They also appear in heterotic 
constructions with $U(1)$ bundles, but for concreteness we center our 
discussion on D-brane models. 

The difficulty in obtaining Majorana masses for the right-handed 
neutrinos is manifest in this setup, since these fields carry non-trivial 
$U(1)$ charges. Typically these $U(1)$ gauge bosons become massive, by 
mixing with a RR closed string modulus, but the symmetries remain as 
global symmetries exact in perturbation theory. Hence it is natural to 
consider the corresponding non-perturbative effects, namely D-brane  
instantons, as the source for the corresponding terms. 

The appearance of non-trivial field theory operators due to 
non-perturbative instanton effects is similar to the appearance of 't 
Hooft operators from gauge theory instantons in theories with mixed $U(1)$ 
anomalies. Namely, the operator arises from path integrating over 
zero modes of the instanton. However, in our setup there are important 
differences with respect to the field theory discussion. First and most 
importantly, $U(1)$ symmetries are actually gauged in string theory
(although as mentioned, it is crucial that the $U(1)$ under which the 
Majorana mass term is charged becomes massive by coupling with a RR 
modulus). This implies that the exponential instanton amplitude in 
\ref{maj} should transform by a phase which cancels the transformation of 
the Majorana mass operator, yielding the instanton amplitude 
gauge-invariant. Secondly, the 
relevant instanton is {\em not} a gauge theory instanton. This has the 
nice consequence that the exponential factor need not lead to a large 
suppression, since it is not related to any SM gauge coupling.

The general description of D-brane instantons, their structure of 
fermionic zero modes, and the effective operators they induce, is carried 
out in complete generality in appendix \ref{generalinst} (in the absence 
of orientifold planes). It also contains the corresponding discussion for 
heterotic models.

In the coming sections we apply this kind of analysis, including 
orientifold planes, to the particular case of generating 
right-handed neutrino Majorana mass terms from string theory 
instantons, in the particular setup of type IIA models of 
intersecting D6-branes. It is however straightforward to rephrase the 
discussion in terms of other D-brane models or of heterotic 
compactifications.

\subsection{Instantons and the right-handed neutrino Majorana mass 
operator}
\label{instantonmech}

As we said,
in order to discuss right-handed neutrino masses we  need to work 
in the context of some semirealistic class of models with  quark/lepton 
generations. For definiteness we are going to present our discussion in 
the context of type IIA orientifolds with D6-branes wrapping 
intersecting 3-cycles \cite{bgkl,afiru} (for reviews see \cite{interev}). 
In this case the relevant instantons are
D2-instantons wrapping 3-cycles on the compact space \footnote{One could 
make an analogous discussion for orientifold compactifications 
of type IIB theory with D$(2p+1)$-branes, like type I models with
magnetized D9-branes \cite{bachas,bgkl,aads}, compactifications with 
D3- and D7-branes \cite{aiq}, models of D-branes at singularities
\cite{aiqu,alday,delta}, or non-geometric constructions like 
orientifolds of Gepner models \cite{gepner2,gepner1,schell}. The 
microscopic description of the 
corresponding D-brane instanton changes, but the physics of the 
four-dimensional theory remains identical. Also, one can make a similar 
discussion in the heterotic side with $U(1)$ bundles.
As discussed in appendix A, in the heterotic the relevant 
operator could be induced simply by  world-sheet instantons 
(hence tree-level in $g_s$ and non-perturbative in $\alpha'$).} 
\cite{Becker:1995kb}.
As mentioned above, the discussion in this section is a particular 
application (including orientifold planes) of the general discussion in 
appendix \ref{generalinst}.

Let us consider stacks of D6-branes $a$, $b$, $c$ and $d$ wrapping 
3-cycles $\Pi_a$, $\Pi_b$, $\Pi_c$ and $\Pi_d$ on the CY orientifold, 
along with their orientifold images, denoted $a^*$, $b^*$, $c^*$, $d^*$ 
branes. Let us denote their multiplicity by $(N_a,N_b,N_c,N_d)$. 
In the literature there are two main classes of semirealistic string 
models, with the chiral content of just the (possibly supersymmetric) 
Standard Model. They differ in the realization of the $SU(2)_L$ gauge factor
of weak interactions, either as coming from a $U(2)$ in two 
overlapping D6-branes away from O6-planes ($N_b=2$) \cite{imr}, or from an 
$USp(2)$ group from one D6-brane ($N_b=1$) overlapping with its orientifold 
image on top of and O6-plane \cite{Cremades:2003qj}. The gauge group will 
be $SU(3)\times SU(2)\times$$U(1)_a\times U(1)_c \times U(1)_d$, with an 
additional Abelian $U(1)_b$ in the first case. This gauge group includes 
that of the SM. 

In order to have the chiral fermion spectrum of the SM one has to 
ensure that the branes intersect the appropriate number of times. 
Thus e.g. left-handed quarks will come from the intersections of $a$ and 
$b$, $b^*$ branes and right-handed quarks from the intersections of $a$ 
and $c$, $c^*$ branes. Right-handed neutrinos will come from 
intersections of $c$ and $d^*$ branes (as discussed above, their charges 
under the $U(1)$ symmetries forbid Majorana mass terms in perturbation 
theory, although can be generated non-perturbatively as discussed below). 
A number of models of this type, with the SM gauge group and chiral 
spectrum, have been constructed in the last few years, and an 
explicit toroidal example will be described in the next section. 
An important point concerning the models we 
focus on is that they have the chiral matter content of exactly the SM. 
This implies that the discussion of global $U(1)$ symmetries and their 
anomalies is exactly as in the Standard model. For reference, the chiral 
fields and their $U(1)$ charges in our models are shown in table 
\ref{tabpssm}.

The $U(1)$ generators $Q_a$, $Q_c$, $Q_d$ have interesting 
interpretations as SM global symmetries. For instance $\frac {1}{3}Q_a$ corresponds 
to baryon number $Q_B$, while $Q_d$ corresponds to (minus) lepton number 
$Q_L$, and $Q_c$ to the $U(1)_R$ generator $Q_R$ of left-right symmetric 
models. Since $U(1)_b$ is not relevant in our discussion (and moreover is 
often absent in 
many interesting models) we do not include it in our discussion
(see appendix \ref{moremodels} for models with $U(1)_b$). Note that 
$U(1)_B$, $U(1)_L$ and $U(1)_R$ are all tree-level symmetries of the SM 
(with right-handed neutrinos). There are three interesting 
orthogonal linear combinations that one can form
\beqa
Q_{\rm anom}\ &  =\ & 3Q_a-Q_d\, =\, 9Q_B+Q_L \nonumber \\
Y\ &=& \ \frac 16 Q_a -\frac 12 Q_c + \frac 12 Q_d \, =\, 
\frac {1}{2}(Q_{B-L}-Q_R)\\
Y'\ & = & \frac 13 Q_a + Q_c + Q_d \, =\, Q_{B-L} + Q_R
\eeqa
The symmetry generated by $Q_{\rm anom}$ is anomalous (with anomaly 
canceled by the Green-Schwarz mechanism), while $Y$ and $Y'$ (equivalently
$U(1)_{B-L}$ and $U(1)_R$) are anomaly free. The generator $Y$ corresponds 
to the standard 
hypercharge, hence it should remain massless in order to have a realistic 
model. Finally, the generator $Y'$ is an extra anomaly-free symmetry, 
to which, by a slight abuse of language, we refer to as the $B-L$ symmetry.
It is crucial for our mechanism to work that this generator, 
even though it is non-anomalous, becomes massive by a St\"uckelberg 
coupling.

As we have argued, Majorana mass terms for right-handed neutrinos are 
forbidden in perturbation theory by the $U(1)_{Y'}$
and $U(1)_{anom}$ symmetries, but can be 
generated non-perturbatively as in (\ref{maj}) by non-perturbative 
instanton effects. Since the scalars making the $U(1)$'s massive are 
obtained from the RR 3-form integrated over 3-cycles,  the relevant 
instantons are euclidean D2-branes wrapped on 3-cycles. Let us consider 
one such instanton corresponding to a D2-brane $M$ (from Majorana)
wrapping a 3-cycle $\Pi_M$ in the compact space, and derive the 
constraints that it must satisfy to lead to operators of the form
\beq
e^{-S_{D2}}\, \nu_R\nu_R
\label{theoperator}
\eeq
Let us postpone for the moment the discussion of how the instanton 
generates the right-handed neutrino Majorana mass operator, and consider 
what are the symmetry properties of such an instanton amplitude.
The right-handed neutrino Majorana mass bilinears $\nu_R\nu_R$ have charge 
$2$ under both $U(1)_{B-L}$ and $U(1)_R$ symmetries, and are neutral 
under hypercharge (and of course baryon number). The corresponding 
transformations under the $U(1)$ gauge symmetries should be canceled by a 
corresponding transformation of the exponential prefactor. This is the 
case if the imaginary part of the instanton action is given by an scalar 
field which shifts under the $U(1)$ symmetry in the appropriate way.
Namely, as already mentioned, for the mechanism to work it is necessary 
that the relevant $U(1)$ gets a St\"uckelberg mass due to a $B\wedge F$ 
coupling to a closed string field.  

Let us consider the general pattern of scalar shifts under the $U(1)$'s. 
Introduce the labels $A,B,\ldots$ for the different brane stacks, and 
their corresponding $U(1)$ gauge symmetries. The 3-cycles $\Pi_A$ on 
which the D6$_A$-brane wrap admit a decomposition in a basis $\{C_r\}$ of 
homology 3-cycles
\beqa
\Pi_A \, =\, \sum_r \, p_{Ar}\, C_r
\eeqa 
As discussed around (\ref{scalarshift}), the coupling of the D6$_B$-branes 
(and their orientifold images $B^*$) to the RR scalars $a_r$ (obtained by 
integrating the RR 3-form over the 3-cycle $D_r$ dual to $C_r$) implies 
that under $A_B\to A_B+d\Lambda_B$, the scalars shift as
\beqa
a_r \, \to \, a_r \, +\, N_B\, (\,p_{Br}-\, p_{B^*r}\,) \Lambda_{B}
\label{shiftorient}
\eeqa
The action of the D2-brane instanton is given by  DBI+ CS action, with 
the latter being responsible for its coupling to the RR scalars. Namely
\beq
\im S_{D2}\ =\ \sum_r c_r\, a_r \ =\ a_M
\eeq
(for supersymmetric D2-branes, the complete action can be expressed 
holomorphically in 
terms of the complex structure moduli). 
Using (\ref{shiftorient}) this quantity shifts under general $U(1)$ 
gauge transformations by the amount
\beqa
-\sum_r \, c_r \sum_A \, N_A\, (\,p_{Ar}-\, p_{A^*r}\,) \Lambda_{A}\, 
=\,-  \sum_{A}\, N_ A\, (\, I_{MA}-I_{MA^*}\,)\, \Lambda_A\, 
\eeqa
where $I_{MA}=\Pi_M\cdot \Pi_A$ is the intersection number, and similarly 
for $I_{MA^*}$. Hence the exponential amplitude for the instanton 
transforms as
\beqa
e^{-S_{D2}} \, \to\, \exp \, (\,- i \, \sum_{A}\, N_ A\, (\, 
I_{MA}-I_{MA^*} \, )\, \Lambda_A\, )\,  e^{S_{D2}} 
\label{cosas}
\eeqa
We thus have that for the exponential factor in (\ref{theoperator}) to 
cancel the transformation of the Majorana mass term, the intersection 
numbers of the D2-brane instanton and the background D6-brane 3-cycles 
must satisfy
\footnote{\label{foot} In addition, if any extra  D6-brane X with $U(n)$ 
group  is present beyond those of the SM, one should also impose 
$I_{MX}-I_{MX^*}=0$ in order for the instanton-induced operator to be 
gauge invariant. See discussion in section \ref{extra}}
\beq
I_{Ma}-I_{Ma^*}= I_{Mb}-I_{Mb^*}\ =\ 0\ \ ;\ \ 
I_{Mc}-I_{Mc^*}=I_{Md}-I_{Md^*}\ =\ 2
\label{condi}
\eeq
Let us now consider the precise microscopic mechanism by which such an 
instanton generates the Majorana mass term insertion. In the presence of 
the D2-brane instanton, there are open strings stretching between the D2- 
and the background D6-branes. Quantization of these open strings 
shows that they lead to chiral fermions localized at the 
intersections between the D2- and the D6-branes. These are fermion zero 
modes of the instanton, over which one should integrate. For 
an instanton with the intersection numbers (\ref{condi}), we find 
two fermionic zero modes of each chirality, denoted $\alpha_i$, 
$\gamma_i$, $i=1,2$ at the intersections of the D2-instanton and the $d$, 
$c$ D6-branes carrying $U(1)_R$ and $U(1)_L$ gauge fields. 

These fermion zero modes have $Mc$-$cd^*$-$d^*$cubic couplings involving 
the D6-D6 fields in the $cd^*$ sector, namely the right-handed neutrino 
multiplets, of the form
\beq
L_{cubic}\ \propto \ d^{ij}_a\ (\alpha_i  \nu^a \gamma_j) \ \ , a=1,2,3
\eeq
The coupling arises via a world-sheet disk amplitude, as shown \footnote{We hope that the 
appearance of two kinds of instantons (the D2-brane 
instanton inducing a non-perturbative $g_s$ correction in the 4d effective 
action, and the world-sheet instanton, inducing an $\alpha'$ effect on the 
D2-brane action) in the present IIA setup does not lead to confusion.
Moreover, the cubic coupling arises as a pure $\alpha'$ tree-level effect 
in other D-brane constructions (namely in type IIB models).} in Figure 
\label{instanton3}.

\begin{figure}
\epsfysize=6cm
\begin{center}
\leavevmode
\epsffile{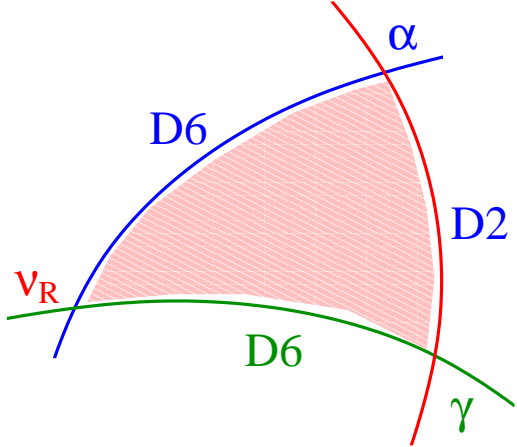}
\end{center}
\caption{World-sheet disk amplitude inducing a cubic coupling on the 
D2-brane instanton action. The cubic coupling involves the
right-handed neutrinos lying at the intersection of the $c$ and $d^*$ 
D6-branes, and the two instanton fermion zero modes $\alpha $ and $\gamma$ 
from the D2-D6 intersections.}
\label{instanton3}
\end{figure}

Upon integration over the fermion zero modes, the complete D2-brane 
instanton amplitude contains the additional contribution
\beq
\int d^2\alpha\,  d^2\gamma \ e^{-d^{ij}_a\ (\alpha_i  \nu^a \gamma_j)}\
\propto \
-\nu_a \nu_b \ \int d^2\alpha\, d^2\gamma
\ \alpha_i\alpha_j\gamma_k\gamma_l\, d^{ik}_ad^{jl}_b
\ =\ \nu_a \nu_b\,  (\, \epsilon _{ij}\epsilon_{kl}d^{ik}_ad^{jl}_b\, )
\quad
\eeq
giving rise to bilinears in the neutrino multiplets in the 4d effective 
action. 
Notice 
the role of the conditions (\ref{condi}) in both pieces of the instanton 
amplitude (\ref{theoperator}). It determines the number of fermion 
zero modes, and their charges, and hence the transformation of the 
monomial in the charged D6-D6 fields. On the other hand, it determines the 
amount by which the exponent $S_{D2}$ shifts. The cancellation between 
the transformations of both pieces is a particular case of the 
self-consistency of these amplitudes, discussed in general in appendix 
\ref{generalinst}.

\medskip

As we have tried to emphasize, the mechanism is rather general, and we 
only need to have a semirealistic compactification with the following 
ingredients: 

1) The 4d theory should have the chiral content of the SM and additional
right-handed neutrinos. There should be a gauged $U(1)_{B-L}$ gauge 
symmetry beyond the SM, under which the right-handed neutrinos are charged. 

2) The $U(1)_{B-L}$ gauge boson should have a St\"uckelberg mass from a 
$B\wedge F$ coupling. 

3) The compact manifold should admit D2-instantons yielding the 
two appropriate zero modes transforming under $U(1)_L$ and $U(1)_R$ (but 
no other symmetries in the theory) to yield neutrino bilinears. 

Then the appropriate Majorana mass term  will generically appear 
(see section \ref{extra} for some additional discussion on more detailed 
conditions on the instantons).

Note that the $e^{-S_{D2}}$ semiclassical factor will provide a 
suppression factor for this operator, but this suppression need not be 
large
\footnote{
One may worry that, if the exponential factor is not small, multiwrapped 
instantons may contribute with comparable strength, leading to a breakdown 
of the instanton expansion. However, the zero mode structure of the 
instanton is controled by the intersection numbers of the overall cycle 
class, thus ensuring that only the single instanton we discuss 
contributes.}, since in 
general the field $U$ is not directly related to the SM gauge coupling 
constants. Indeed, it is easy to see that $U$ cannot appear in the gauge 
kinetic function for the SM gauge fields. The reason is that $U$ 
transforms with a shift under an anomaly free $U(1)$. If it also had 
couplings to the $F\wedge F$ SM gauge field operators a gauge anomaly 
would be created, which cannot be true for an anomaly-free gauge 
interaction. Thus $U$ cannot appear in the SM gauge kinetic functions and 
hence there is no phenomenological constraint on the value of $ReU$. Thus 
the induced Majorana mass for right-handed neutrinos may be only a few 
order of magnitudes below the string scale, in agreement with 
phenomenological requirements. Note that in general a flavour structure 
will appear depending on the model dependent coefficients $d^{ij}_a$.

An important comment concerning discrete symmetries is in order.
The mentioned instanton effects leading to the operator 
(\ref{theoperator}) break the $U(1)_{B-L}$ continuous symmetry. 
Notice however that a $Z_2$ group generated by $exp(i\pi Q_{B-L})$ remains 
unbroken (i.e. exp(-U) has lepton charge -2). On the other hand it is
well known that within the MSSM such a discrete $Z_2$ symmetry
is equivalent to R-parity \cite{ir}, the symmetry which guarantees
the absence of dimension four operators violating Baryon and
Lepton numbers in the MSSM. Then  within the present scheme  the
existence  of R-parity is automatic.

\medskip

The present mechanism may be implemented in a way consistent with $SU(5)$ 
unification. The idea is having a $SU(5)\times U(1)_{Z}$ model with three
chiral matter generations including 3 right-handed neutrinos, i.e.
\beq
 3(10_{1}+{\bar 5}_{-3}+1_5)
\eeq
It is easy to check that the $U(1)_Z$ is anomaly-free generation
by generation \footnote{In fact one has $U(1)_Z=U(1)_Y+(5/2)U(1)_{Y'}$, 
with $U(1)_{Y'}$ the massive anomaly-free $U(1)$. Note also that the 
charge assignments are compatible with embedding the $SU(5)\times U(1)_Z$ 
into $SO(10)$. However, this enhancement would not be consistent with our 
mechanism, which requires the existence of a St\"uckelberg mass term for 
the 
relevant $U(1)$. Hence, after the instanton effect is taken into account, 
only the $SU(5)\times Z_2$ symmetry would be realized at low energies,
with $Z_2$ being R-parity in the SUSY case.}. An instanton with an action 
whose imaginary part $X$ transforms under $U(1)_Z$ like
\beq
X\ \longrightarrow \ X \ +\ 10 \Lambda_{Z}
\eeq
would generate the operator $e^{-X}\nu_R^i\nu_R^j$, which is invariant 
under the gauge symmetry. It would be interesting to have some concrete 
$SU(5)$ example within string theory  where this could be implemented.

\subsection{Role of supersymmetry and  additional zero modes}
\label{extra}

In our previous discussion we have focused on the relevant properties of 
the instanton to yield the effect we are interested in. These are 
essentially based on symmetry arguments, and topological properties. In 
particular, the previous analysis ignores the discussion of other 
features, like the role of supersymmetry, the possible presence of 
additional instanton zero modes, etc. This section is devoted to filling 
this gap.

\medskip

{\bf The role of supersymmetry}

As with most physical effects in string theory compactifications, 
instanton corrections have usually been described in the supersymmetric setup.
Indeed, beyond the usual advantages of ensuring stability of the vacuum, 
and that the wrapped brane is volume minimizing and thus a stationary 
point of the path integral, $N=1$ supersymmetry provides a useful 
bookkeeping which facilitates 
the classification of the spacetime operators induced by the instanton. For 
instance, instanton corrections to the spacetime superpotential arise from 
instantons with two fermionic zero modes, which soak up the integration 
over half the superspace Grassman variables. These are usually generated 
by D-brane or world-sheet instantons, wrapped on rigid cycles, 
and preserving half of the supersymmetries. This ensures that the only 
zero modes are the Goldstinos of the two supersymmetries broken by the 
instanton, so that it generates a superpotential coupling.

However, it is clear that instanton effects exist in non-supersymmetric 
theories as well. Essentially the basic rule is that an instanton with 
a number of fermion zero modes leads to spacetime interactions with the 
appropriate number of spacetime fields to saturate the amplitude. Given 
this, we 
understand that our previous discussion of generation of Majorana mass 
terms from instantons can be carried out both in supersymmetric and 
non-supersymmetric string compactifications. In fact, our explicit 
examples in the coming sections are non-supersymmetric. In any event, it 
should be clear that a completely analogous analysis can be carried out 
for supersymmetric compactifications.

\medskip

{\bf Additional zero modes}

A physically more relevant issue is that in general an instanton may carry 
more zero modes than the minimum we require. More specifically, one may 
have instantons with the right topological intersection numbers, but with 
additional zero modes associated to deformations of the wrapped 3-cycle, 
(of for instantons that happen to break all the supersymmetries of a 
supersymmetric background, and hence have additional Goldstinos).
 These zero modes are uncharged under the D6-brane gauge factors, and 
therefore do not contribute to the structure of the spacetime operator in 
the charged fields (but rather to insertions of additional closed string 
fields). Since these zero modes do not really affect the Majorana mass 
term structure, their detailed discussion is ignored in the present paper.

Such additional zero modes will however be present for the instantons we 
consider 
in our explicit examples. Indeed we present explicit examples in 
toroidal setups, where the 3-cycles wrapped by the D2-brane instantons 
are topologically $T^3$, hence have three position plus Wilson line 
moduli, leading to additional zero modes. This may be considered a 
drawback, since as discussed leads to additional closed string field 
insertions, making the Majorana mass term structure be part of a 
higher-dimension operator (in particular, it would not be a 
superpotential coupling in supersymmetric cases). These models however 
illustrate the robust features in the generation of Majorana mass terms, 
and can be considered toy models of more realistic constructions (e.g. on 
CY threefolds) where instantons without the additional zero modes may 
exist.

Moreover, even the discussion of the toroidal setups may be useful by 
itself, in the following sense. It is well-known that additional 
bosonic and non-chiral fermionic instanton zero modes can be lifted by 
additional ingredients in 
the compactification. For instance, 3-form fluxes on type IIB orientifold 
models can lift certain fermion zero modes on euclidean D3-brane 
instantons \cite{fluxinst} (and consequently, modify the topological 
conditions for an instanton to contribute to the superpotential). It is 
plausible to imagine that the additional zero modes in the models we 
discuss can be lifted by a similar mechanism upon introduction of a 
suitable set of fluxes, or generalization thereof (see 
\cite{Marchesano:2006ns} for a useful related discussion in the type IIA 
setup). Notice that in non-supersymmetric cases, this removal of the additional 
zero modes also involves stabilization of e.g. the deformation 
moduli of the wrapped cycle. This effect can in fact be crucial, since 
it underlies that fact that the wrapped brane is an stationary point of 
the path integral, and can thus be properly refered to as an instanton.

An additional possible source of additional zero modes is that 
the D2-brane may intersect other D6-branes beyond those involved in the SM 
sector. This possibility is fully encompassed by our general discussion in 
appendix \ref{generalinst}. Such additional zero modes would lead to 
insertions of new D6-D6 fields in the spacetime effective operator, thus 
spoiling in principle the Majorana mass term structure, and in general 
yielding a higher-dimension operator, involving hidden sector fields. 
Nevertheless, the right structure may be recovered if these new fields 
are allowed to get vevs, etc. Clearly the discussion then becomes very 
model-dependent. In our explicit neutrino mass models we will make sure that these 
additional zero modes are absent (see footnote \ref{foot}).

There is one interesting exception to this last paragraph. Certain 
instantons contain zero modes arising from equal numbers of intersections 
between the D2-brane with a given D6-brane $A$ and its orientifold image 
$A^*$, namely $I_{MA}-I_{MA^*}=0$. This leads to zero modes in the 
$(\fund_M,\antifund_A)+(\fund_M,\fund_A)$, and are hence vector-like with 
respect to the $U(N_A)$ gauge symmetry. From the discussion in 
previous sections, see e.g. \ref{cosas} these zero modes do not affect the 
transformation of the exponential factor. Consistently with this, they do 
not have any cubic couplings with 4d fields, hence do not contribute 
insertions to the charged matter operator. These intersection zero modes 
are thus particularly inert, and we do not consider them in our 
discussion.

\medskip

{\bf Moduli stabilization}

We would like to conclude this section with a comment on the interplay of 
the wrapped brane instantons and the introduction of additional 
ingredients in the compactification, like fluxes. Indeed, since our 
mechanism involves the use of a shift for the imaginary part of a closed 
string modulus, one may fear that it is spoiled by the introduction of 
fluxes leading to closed string stabilization. In fact, this is guaranteed 
{\em not} to be the case. As discussed in \cite{Camara:2005dc} the 
Freed-Witten constraint \cite{Freed:1999vc} on the D6-branes guarantees 
that the superpotentials generated by the introduction of fluxes are fully 
compatible with the shift symmetries induced by the $BF$ couplings due to 
the D6-branes. In other words, the modulus involved in the instanton 
amplitude is not affected by the flux superpotential, and the shift 
symmetry is intact.

Amusingly the converse result was shown in \cite{Kashani-Poor:2005si}. 
Namely, the Freed-Witten constraint on the D2-brane instanton guarantees 
that the superpotential induced by such instantons is compatible with the 
scalar shifts implicitly exploited by the fluxes (manifest in their 
description as gaugings of isometries of the scalar moduli space).

The bottomline is that flux stabilization mechanisms and instantons talk 
to different sets of moduli, and hence lead to no interference. This gives 
additional plausibility to our statements above concerning lifting of 
additional instanton zero modes, and hence motivates us to proceed with 
the construction of explicit models, even in toroidal setups.

\section{An intersecting D6-brane example}
\label{example}

To make the above described general mechanism explicit, we need 
semirealistic models (either supersymmetric or not) in which there is a 
gauged $U(1)_{B-L}$ symmetry getting a St\"uckelberg mass. These are a 
restricted subset of the semirealistic models in the literature.
To our knowledge, the only models satisfying those requirements are
the non-SUSY models constructed in \cite{imr}, and a (small) subset of 
the SUSY CFT orientifolds studied by Schellekens and collaborators
\cite{schell}. In particular, there are no examples of toroidal/orbifold  
$N=1$ SUSY constructions with massive $U(1)_{B-L}$ gauge bosons. In any 
event, and given the simplicity of toroidal constructions, our examples 
here will be analogous to the non-SUSY models in \cite{imr}. 
As just explained supersymmetry is not a crucial ingredient in our 
discussion, the relevant instantons and 't Hooft operators also exist 
in non-supersymetric theories. In our case the relevant operator will be 
fermionic with a bilinear in right-handed neutrinos. It should be 
easy to implement a similar discussion in the supersymmetric models in 
\cite{schell}. 

As mentioned, ref.\cite{imr} provided a family of non-SUSY models with 
the required characteristics, i.e. SM spectrum and a massive $U(1)_{B-L}$.
They are orientifolds of type IIA on $T^2\times T^2\times T^2$ modded by 
$\Omega R$, with $\Omega $ being world-sheet parity and $R$ the reflection  
$z_i\to \ov z_i$ of the three $T^2$ complex coordinates.
There are four Standard Model D6-branes  a,b,c,d (and their orientifold 
images $a^*,b^*,c^*, d^*$) in which the SM gauge group lives.
The multiplicities are $N_a=3$, $N_b=2$, $N_c=N_d=1$ so that, before
some gauge bosons get St\"uckelberg masses, the full gauge group is
$U(3)_a\times U(2)_b\times U(1)_c\times U(1)_d$. In the present section we 
will consider a slightly simpler class of models \cite{cimunpub} in which 
the $SU(2)_L$ SM gauge group is realized in terms of a symplectic group
$USp(2)$ rather than a unitary group $U(2)$ \footnote{See appendix B for 
the analogous discussion for the constructions in \cite{imr} which have a 
$U(2)_b$ gauge group.}. This is obtained with $N_b=1$ if the corresponding 
b-brane and its mirror sit on top of an orientifold plane. Then the 
initial gauge group is rather $SU(3)\times SU(2)\times U(1)_a\times 
U(1)_c\times U(1)_d$. Here $U(1)_a$ and $U(1)_d$ have the interpretation 
of gauged baryon and (minus)lepton numbers, whereas $U(1)_c$ behaves like 
the diagonal generator of right-handed weak isospin. Open strings at the 
intersections of the D6-branes lead to chiral fermions transforming like 
bifundamentals $(\fund_a,\antifund_b)$, and $(\fund_a,\fund_b)$ for $ab$ 
and $ab^*$ intersections, respectively. The chiral fermion content 
reproduces the SM quarks and leptons if the D6-brane intersection numbers 
are given by
\beqa
I_{ab}=I_{ab^*}= \ 3 \ \ & ; &\ \ I_{ac}=I_{ac^*}= \ -3 \nonumber \\
I_{db}=I_{db^*}= \ -3 \ \ & ;  & I_{cd}=\ -3 \ ;\ I_{cd^*}=\ 3 
\label{intersm}
\eeqa
with the remaining intersections vanishing. As usual, negative 
intersection 
numbers denote positive multiplicities of the conjugate representation.
The spectrum of chiral fermions is shown in Table \ref{tabpssm}.
%
\begin{table}[htb] \footnotesize
\renewcommand{\arraystretch}{1.25}
\begin{center}
\begin{tabular}{|c|c|c|c|c|c|c|c|c|}
\hline Intersection &
 Matter fields  &   &  $Q_a$  &   $Q_c $ & $Q_d$  & Y & Y'&$3Q_a-Q_d$
\\
\hline\hline (ab),(ab*) & $Q_L$ &  $3(3,2)$ & 1  & 0 & 0 & 1/6 & 1/3 & 3 \\
\hline (ac) & $U_R$   &  $3( {\bar 3},1)$ &  -1   & 1  & 0 & -2/3& 2/3 & -3
\\
\hline (ac*) & $D_R$   &  $3( {\bar 3},1)$ &  -1    & -1  & 0 & 1/3 &-4/3&-3
\\
\hline (bd),(b*d) & $ L$    &  $3(1,2)$ &  0    & 0  & -1 & -1/2 & -1 & 1\\
\hline (cd) & $E_R$   &  $3(1,1)$ &  0   & -1  & 1  & 1 & 0 & -1\\
\hline (cd*) & $\nu_R$   &  $3(1,1)$ &  0   & 1  & 1  & 0 & 2 & -1 \\
\hline \end{tabular}
\end{center} \caption{ Standard model spectrum and $U(1)$ charges in the 
realization in terms of D6-branes with intersection number (\ref{intersm})
\label{tabpssm} }
\end{table}
%
These correspond to three SM quark lepton generations. In addition there are 
three right-handed neutrinos $\nu_R$ whose presence is generic in this kind
of constructions. At the intersections there are also complex scalar with 
the same charges as the chiral fermions. These are not necessarily 
massless (since the model may be non-supersymmetric), but by a judicious 
choice of the complex structure moduli one can generically avoid the 
presence of charged scalar tachyons \cite{imr}.

One linear combination of the three $U(1)$'s, i.e. 
\beq
Y \ =\ \frac {1}{6}\left( Q_a\ -\ 3Q_c\ +\ 3Q_d\right)
\eeq
corresponds to the hypercharge generator. Another one, $(3Q_a-Q_d)$ is 
anomalous (with anomaly canceled by the Green-Schwarz mechanism) and 
becomes massive as usual. The remaining orthogonal linear combination $Y'$ 
is anomaly free and will become massive or not depending on the structure 
of the couplings of the $U(1)$'s to the RR 2-forms in the given model. As 
we mentioned, the appearance of a Majorana mass term by our mechanism 
necessarily requires that this anomaly-free combination becomes massive, 
otherwise the term is forbidden by unbroken gauge interactions. 
As discussed in the previous section, in order for such mass terms to be 
generated there must exist a D2-brane instanton $M$ with intersection 
numbers with the SM branes as in eq.(\ref{condi}). Taking into account 
that the helicities of the $\alpha$ and $\gamma $ instanton zero modes 
have to match, this requires either
\beq
I_{Md}=2\ ;\ I_{Mc*}=-2 \ ;\ I_{Mc}=I_{Md*}= 0
\label{condiM1}
\eeq
or else
\beq
I_{Mc}=2\ ;\ I_{Md*}=-2 \ ;\ I_{Md}=I_{Mc*}= 0 \ .
\label{condiM2}
\eeq
Thus in order to get a model with the SM chiral content and in addition 
right-handed Majorana masses both the conditions (\ref{intersm}) and those 
above must be verified. The latter conditions turn out to be rather 
restrictive in the present setup, and we suspect this to be valid in more 
general classes of models. 

Let us make the discussion concrete and construct a specific class of 
models. Consider type IIA string theory compactified on $T^2\times 
T^2\times T^2$, with $(x_i,y_i)$ parametrizing the $i^{th}$ $T^2$. We 
further mod out by $\Omega R$, where $\Omega$ is world-sheet parity and 
$R$ is the reflection of the three compact $y_i$ coordinates.   
We consider D6-branes on factorizable 3-cycles and denote their wrapping 
numbers in the three $(x_i,y_i)$ directions $(n^1,m^1)$, $(n^2,m^2)$, 
$(n^3,m^3)$. Consider a set of SM branes with wrapping numbers 
as shown in Table (\ref{newfamily}). 
%
\begin{table}[htb] \footnotesize
\renewcommand{\arraystretch}{2.5}
\begin{center}
\begin{tabular}{|c||c|c|c|}
\hline
 $N_i$    &  $(n_i^1,m_i^1)$  &  $(n_i^2,m_i^2)$   & $(n_i^3,m_i^3)$ \\
\hline\hline $N_a=3$ & $(1,0)$  &  $(n_a^2, 1)$ &
 $(n_g,  m_a)$  \\
\hline $N_b=1$ &   $(0,1)$    &  $ (1,0)$  &
$(0,-1)$   \\
\hline $N_c=1$ & $(n_c^1,1)$  &
 $(1,0)$  & $(0,1)$  \\
\hline $N_d=1$ &   $(1,0)$    &  $(n_d^2,-n_g)$  &
$(1, m_d^3)$   \\
\hline \end{tabular}
\end{center} \caption{ D6-brane wrapping numbers giving rise to a SM spectrum.
\label{newfamily} }
\end{table}
Here $n_a^2$, $m_a^3$, $n_c^1$, $n_d^2$, $m_d^3$ are integers. 
It is easy to check that indeed these wrapping numbers give rise to the 
chiral spectrum of a SM with $n_g$ quark/lepton generations. For more 
generality we have considered the case with a general number of 
generations $n_g$. 

These  models have in principle three 
$U(1)$ gauge fields. However generically two of them acquire
St\"uckelberg masses  due to the $B\wedge F$ couplings
\beqa
 B_2^i \wedge 2 N_A m_A^i n_A^j n_A^k F_A \quad , \quad 
i\neq j\neq k
\eeqa
where $A$ labels the D6-brane stacks and $i,j,k$ run through the three 
2-tori. The factor of $N_A$ arises from the $U(1)$ normalization, and the
factor of 2 arises from the coupling to the D6-brane and its orientifold
image. Recall that in these toroidal models there are four massless 
2-forms $B_2^p$, $p=0,1,2,3$ arising from integrating the type IIA 5-form
over 3-cycles invariant under the orientifold action. For the model in 
table \ref{newfamily} the non-vanishing couplings are

\beqa
B_2^2 &\wedge  & \ 2n_g(3F^a\ -\  F^{d})
\nonumber \\
 B_2^3 &\wedge  &  \ 
2(\,  3n_a^2m_a^3 F^a\ +\ 
 n_c^1 F^c \ +\ n_d^2m_d^3 F^d\, ) \
\label{bfs}
\eeqa
The RR fields $B_2^0$ and $B_2^1$ have no couplings to the $U(1)$'s.
The existence of these couplings implies that  the scalar fields $a_i$,
4d duals to the 2-forms $B_i$, transform under $U(1)$ gauge transformations 
with a shift
\beqa
a_{0,1}\ &\rightarrow &\ a_{0,1} \nonumber \\
a_2\ &\rightarrow & \ a_2\  + \ 2n_g
(3\Lambda(x)_{a}\ -\ \Lambda(x)_d)\\
a_3\ &\rightarrow & \ a_3\  + 6n_a^2m_a^3 \Lambda_a(x)
\ +\ 2n_c^1 \Lambda_c(x) \ +\ 2n_d^2m_d^3 \Lambda_d(x) \nonumber
\label{transfor}
\eeqa
Note that the $a^i$ are the imaginary parts of the complex
structure fields $U^i$ in the case of SUSY  type IIA orientifolds.
There are a number of additional constraints to make the model realistic 
and to allow the non-perturbative appearance of right-handed 
neutrino Majorana masses:

{\it i)}
There are three $U(1)$'s and only two RR-scalars, $a^2$ and $a^3$, 
coupling to them. Hence necessarily one of the $U(1)$'s remains massless.
In order for the model to be realistic, the standard hypercharge should be 
the massless generator. This is the case if
\beq
n_c^1\ =\ n_a^2m_a^3+n_d^2m_d^3 \ .
\eeq
The other two linear combinations (in particular the $U(1)$ relevant 
for Majorana masses) are massive.

{\it ii)}
In order for the model to be a consistent compactification, RR-tadpoles 
have to cancel. Tadpoles cancel in this simple model if
\beq
3m_a^3\ =\ n_g m_d^3 \ \ .
\eeq
In addition one should add $(3n_a^2n_g+n_d^2-16)$ D6-branes (or 
antibranes, depending on the sign) along the orientifold plane. They
have no intersection with the rest of the branes and do not modify 
the discussion in any way.

{\it iii)}
Finally, the appearance of Majorana masses requires the existence of
a D2-brane instanton $M$ wrapping a 3-cycle on $T^2\times T^2\times T^2$ 
verifying the conditions (\ref{condiM1}) (conditions \ref{condiM2} lead 
to equivalent physics). For simplicity we focus on factorizable 3-cycles
(see section \ref{extra} for discussion on our viewpoint on the extra 
zero modes that arise).
It turns out that in this class of models there is a unique factorizable 
3-cycle with the required properties which is given by the wrapping numbers:
\beq
M\ =\ (n_1,m_1)\,(n_2,m_2)\,(n_3,m_3)\ =\ (1,1)\,
(\,\frac {n_a^2}{n_g}, 1)\,(1,-m_a^3)
\label{instatan}
\eeq
In addition in order to get (integer) quantized wrapping numbers for the 
instanton $M$, one must have  $n_c^1=1$ and $n_a^2$ a multiple of $n_g$. 

It is amusing that one can check  that in this particular family of 
models all these three conditions i)-iii) are possible only for $n_g=3$, 
i.e. only for three quark-lepton generations. This is mostly due to the 
condition from cancellation of RR-tadpoles. Although this probably will 
not 
be the case for other classes of models, it illustrates how restrictive 
the conditions to get Majorana masses may be in particular families of 
models. We expect this general lesson to extend to other classes of 
models. Another example of this is the number of Higgs multiplets. Pairs 
of Higgs doublets arise from open strings stretched between the $b$ and 
$c$ D6-branes, and the number of pairs is given by the intersection number 
of the two stack in the last two complex planes, namely $n_c^1$ in our 
class. Since the above conditions (in particular wrapping numbers for the 
instanton 3-cycle) required having $n_c^1=1$, they imply the requirement 
of having just one pair of Higgs doublets.

Let us verify that the above instanton has the correct transformation 
properties. The imaginary part of the action of the D2-brane instanton 
wrapping this 3-cycle is
\beq
\im S_{D2} \ =\ \frac {n_a^2}{n_g}\, a_0\ -\ m_a^3\, a_1\ -\ 
\frac {n_a^2m_a^3}{n_g}\, a_2\ +\ a_3
\label{accionins}
\eeq 
Now it is straightforward to check that the operator
\beq
\exp\ [ \, -\, (\, \frac {n_a^2}{n_g} a_0
-m_a^3 a_1 - \frac {n_a^2² m_a^3}{n_g}a_2
+ a_3\, ) \, ]\, \nu_R^i\nu_R^j \ \ ; \ \ i,j=1,2,3
\label{finaloperator}
\eeq
is gauge invariant under all three $U(1)$'s, when 
one takes into account the transformations (\ref{transfor}).

As discussed in the previous section, the Majorana mass operator is
generated as follows. At the intersections between the D2-brane instanton 
3-cycle with the background D6-branes there are fermionic zero modes.
Let us the denote these  modes as $\alpha_i$, $\gamma_i$
for the $Mc*$ and $Md$ intersections respectively. These zero modes have 
cubic couplings
\beq
L_{cubic}\ \propto \ d^{ij}_a\ (\alpha_i  \nu^a \gamma^j) \ \ , a=1,2,3
\eeq
which are induced by disk world-sheet instantons. Here $\nu^a$ are the 
right-handed sneutrinos and $d^{ij}_a$ are coefficients (which in general 
will also depend on the K\"ahler moduli and open string 
moduli, like standard Yukawas). Upon integration over the fermion zero 
modes, one generates a contribution proportional to
\beq
\int d^2\alpha \, d^2\gamma \ e^{-d^{ij}_a\ (\alpha_i  \nu^a \gamma_j)}\
\propto \
-\nu_a \nu_b \ \int d^2\alpha \, d^2\gamma
\ \alpha_i\alpha_j\gamma_k\gamma_l\, d^{ik}_ad^{jl}_b
\ =\ \nu_a \nu_b\, (\,\epsilon_{ij}\epsilon_{kl}d^{ik}_ad^{jl}_b\, )
\quad
\label{formulita}
\eeq
Note that we get bilinears because we have two zero modes of each type 
$\alpha $ and $\gamma$.

The semiclassical contribution to the quantities $d^{ij}_a$ may be 
explicitly computed in these toroidal models. Indeed 
these amplitudes are completely analogous to the Yukawa
couplings computed in \cite{Cremades:2003qj}. In each of the subtori 
the branes $c^*$, $d$ and the instanton $M$ intersect forming
triangles. Being in a torus we have in fact a sum over 
triangles in the covering space. This computation was performed 
in \cite{Cremades:2003qj} and it was found that the amplitudes may be 
written as products of Jacobi $\theta $-functions with 
characteristics. In particular one finds 
\beq
d^{ij}_a = 
\prod_{r = 1}^3
\vt \left[
\begin{array}{c}
\d^{(r)} \\ \phi^{(r)}
\end{array}
\right] (\k^{(r)}),
\label{totalyuki2}
\eeq
where the product goes over the three tori.
The dependence on $i,j,a$ is contained in the arguments which are
defined as
\beqa
\d^{(r)} & = & \frac{i^{(r)}}{I_{Mc^*}^{(r)}} 
+ \frac{j^{(r)}}{I_{dM}^{(r)}} 
+ \frac{a^{(r)}}{I_{c^*d}^{(r)}} 
+ \frac{  \left(I_{Mc^*}^{(r)} \eps_d^{(r)} 
+ I_{dM}^{(r)} \eps_{c^*}^{(r)} + I_{c^*d}^{(r)} \eps_{M}^{(r)}\right)}
{I_{Mc^*}^{(r)} I_{dM}^{(r)} I_{c^*d}^{(r)}} ,
\label{paramT2ncpx1} \nonumber\\
\phi^{(r)} & = & 
\left(I_{Mc^*}^{(r)} \th_d^{(r)} + 
I_{dM}^{(r)} \th_{c^*}^{(r)} + 
I_{c^*d}^{(r)} \th_{M}^{(r)}\right) , 
\label{paramT2ncpx2}\\
\k^{(r)} & = & 
\frac{J^{(r)}}{\a^\prime} 
|I_{Mc^*}^{(r)} I_{dM}^{(r)} I_{c^*d}^{(r)}|
\label{paramT2ncpx3} \nonumber
\eeqa
Here e.g. $I_{Mc^*}^{(r)}$ denotes the intersection
number of branes $M$ and $c^*$ in the $r$-th torus, and
$i^{(r)}$ labels one  particular intersection in each plane $r$.
Note that in our case 
\beq
I_{Mc^*}^{(1)}I_{Mc^*}^{(2)}I_{Mc^*}^{(3)}=
I_{dM}^{(1)}I_{dM}^{(2)}I_{dM}^{(3)}=\ 2 \
\eeq
\beq
I_{c^*d}^{(1)}I_{c^*d}^{(2)}I_{c^*d}^{(3)}= \ 3  \ .
\eeq
The $J^{(r)}$ are the complex K\"ahler moduli for each tori,
the $\epsilon ^{(r)}$'s parametrize the position of each brane
in each subtorus and the $\theta ^{(r)}$ possible Wilson lines.
These two degrees of freedom correspond to open string moduli 
which may be complexified as
\beq
\Phi_a^{(r)} = J \eps_a^{(r)} + \th_a^{(r)}.
\label{omoduli}
\eeq
As discussed in section \ref{extra}, the D2-brane moduli really correspond 
to instanton zero modes over which one should in principle integrate. 
However, our viewpoint is that the present model should be regarded either 
as a toy model of an improved setup, like CY compactifications, where 
instantons without such zero modes exist, or as part of a construction 
with additional ingredients, like fluxes, which lift such zero modes. 
Either viewpoint is essentially mimicked by considering the D2-brane 
moduli as fixed numbers in the above formulae.

\begin{figure}
\epsfysize=6.5cm
\begin{center}
\leavevmode
\epsffile{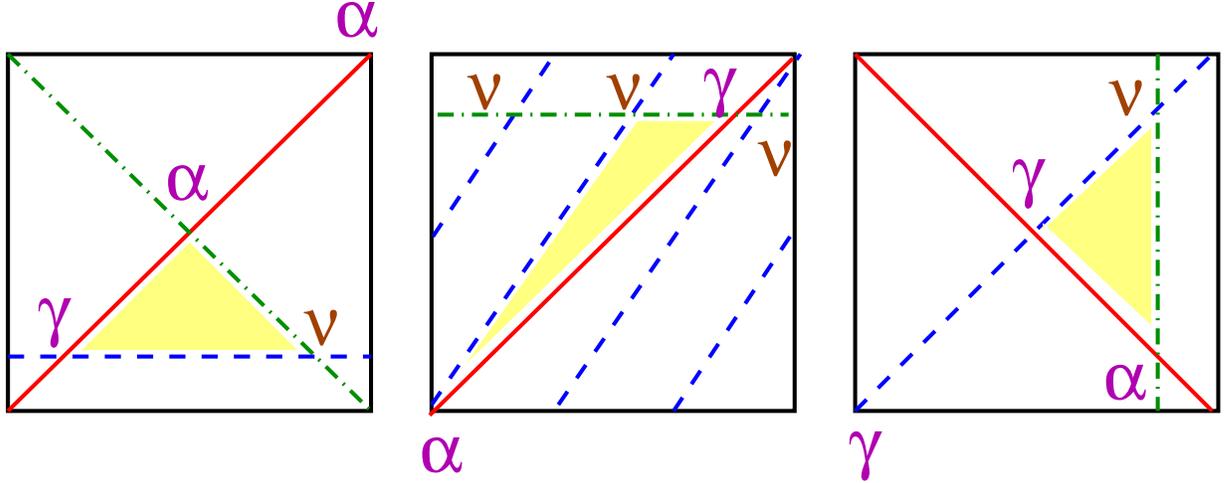}
\end{center}
\caption{The figure shows the D2-brane instanton (continuous line) and 
the $c^*$
and $d$ D6-branes at whose intersections lie the right-handed neutrinos.
The instanton zero modes from the D2-D6 open strings are denoted by
$\alpha$ and $\gamma$. The yellow areas describe (the projections of) the
open string disk inducing a cubic coupling on the D2-brane instanton
action.}
\label{instantex}
\end{figure}

Let us provide a geometric picture of the instanton for a particular example.
Consider the case
\beq
n_a^2=3\ ; n_c^1=1\ ; n_d^2=-2\ ; \ m_a^3=m_d^3=1 \ ;\ n_g=3
\ ; \epsilon=1
\eeq
Then the relevant instanton $M$ and branes $c^*$ and $d$ have
wrapping numbers:
\beqa
M &:& (1,1)(1,1)(1,-1) \nonumber \\
c^*&:& (1,-1)(1,0)(0,-1)\\
d &:& (1,0)(-2,-3)(1,1) \nonumber
\eeqa
Then one can check $I_{dM^*}=I_{Mc}=0$ so that there are no vector-like
zero modes from extra intersections. These three 3-cycles are shown in  
figure \ref{instantex}.
Note that they have the correct number of intersections and also that 
the expected triangle instanton contributions are indeed present.

The models discussed in this section come remarkably close to many of the 
features of the SM. On the other hand they are  not fully realistic.
In fact if the torus is factorizable (no off-diagonal Kahler moduli )Ã
the index dependence of the $d^{ij}_a$ factorizes 
(i.e. $d^{ij}_a \ =\ d^id_ad^j$) and the amplitude vanishes
due to the  contraction with antisymmetric indices in eq.(\ref{formulita}).
Thus only for non-diagonal Kahler moduli the mechanism 
may take place.
Furthermore,  being non-supersymmetric
one expects the vacuum to get unstable unless the models are supplemented 
with some extra ingredient like RR/NS fluxes. Still we think they 
exemplify in very explicit detail how our proposed mechanism 
for the generation of right-handed neutrino Majorana masses works.
We leave for the future the detailed study of the  flavour
patterns which may arise in this class of models.

\section{Other instanton induced superpotentials. The $\mu$-term.}
\label{muterm}

\begin{figure}
\epsfysize=5.5cm
\begin{center}
\leavevmode
\epsffile{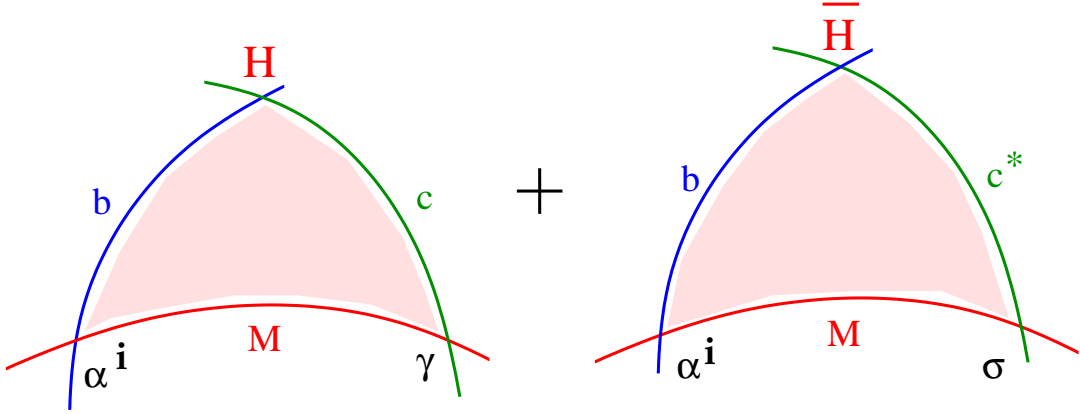}
\end{center}
\caption{Disk amplitudes contributing cubic couplings between the 
D2-brane
instanton fermion zero modes $\alpha^i$, $\gamma$, $\sigma$, and the
spacetime Higgs fields $H$, $\ov H$. Upon integration over the
fermion zero modes, the induced effective operator is a $\mu$-term.
}
\label{instanmu}
\end{figure}

It is clear that this kind of instanton effects may give rise to other 
interaction terms. This includes mass terms as well as certain higher 
dimensional superpotential couplings.  Potentially the most relevant terms 
are those of dimension smaller than four like the case of right-handed 
neutrino masses just mentioned. In the context of the MSSM the only
other (R-parity preserving)  
mass term which is allowed by SM gauge symmetries is a
Higgs bilinear superpotential, i.e. the $\mu $-term
\beq
W\ =\ \mu \ H{\bar H}  \ \ .
\eeq
One of the mysteries  of the MSSM is the understanding of the reason why 
such a mass term, which in principle could be as large as the Planck 
scale, is so small, of the same order of magnitude of the electroweak 
scale. This is often called `the $\mu$-problem'
\cite{kn}.
A natural idea is to assume that such a coupling is forbidden by some 
$U(1)$ gauge interaction. If such a a $U(1)$ gets a St\"uckelberg mass, 
then 
an operator of the general form
\beq
W_{\mu } \ =\ e^{-S_{ins}}H{\bar H}M_{string}
\eeq
may be gauge invariant and be induced by some string instanton contribution.
The exponential suppression could then perhaps provide a dynamical 
explanation for the smallness of the $\mu $-term. The general idea can be 
described without need of a specific model. We consider again the case of 
a general Type IIA CY orientifold with intersecting D6-branes, although 
again a similar discussion can be carried out in other string 
constructions. Consider the case of four stacks of SM branes $a$, $b$, 
$c$, $d$ (and their mirrors) leading to a general unitary group 
$U(3)_a\times U(2)_b\times U(1)_c\times U(1)_d$. The Higgs fields $H$ and 
${\bar H}$ will appear at the $bc$ and $bc^*$ intersections, respectively. 
The bilinear $H{\bar H}$ has $U(1)_b$ charge $\pm 2$, depending on the 
sign of the intersection number of both branes. Then the $\mu $-term 
operator explicitly breaks $U(1)_b$ gauge invariance. In general $U(1)_b$ 
is anomalous and gets a St\"uckelberg mass as usual. Let us assume for 
definiteness that the $U(1)_b$ charge of the bilinear is $-2$, and also 
that both $H$ and ${\bar H}$ come only in one copy, as in the MSSM. Then, 
if a D2-instanton $M$ exists such that
\beqa
\ I_{Mb}\ =\ -1  \ \  I_{Mb*}\ =\ 0\ \ ;\ \  I_{Mc}\ =\ I_{Mc*}\ =\ 1 \ \
;\ \ I_{Mx}\ -\ I_{Mx*}\ =\ 0
\label{condmu}
\eeqa
with $x$ any other brane in the model, then an operator  of the required 
form appears. This means that there should be a doublet $\alpha^i$ of
$Mb$ zero modes, and in addition singlet zero modes $\gamma$, $\sigma$
corresponding to $Mc$ and $Mc*$ intersections. There is a cubic coupling
of these modes to the spacetime Higgs fields, of the form
\beq
L_{cubic} \ \propto\ (\alpha H)\gamma \ +\ (\alpha {\bar H})\sigma
\eeq
This is mediated by the world-sheet disk instanton amplitudes depicted in
fig.(\ref{instanmu}).
%
%

Integration over the fermions zero modes gives a contribution proportional 
to
\beq
e^{-S_{M}}\int d^2\alpha d\gamma d\sigma\ e^{(\alpha H)\gamma + (\alpha 
{\bar H})\sigma  }\
\propto \  e^{-S_{M}}
H{\bar H}
\eeq
We will illustrate this possibility in an explicit MSSM-like example in
Appendix C. It turns out that if we want to construct an explicit
RR-tadpole free SUSY model, extra new D6-branes  beyond the SM ones
have to be introduced. Then the actual operator gets multiplied by
a power of hidden sector fields. Still it provides an explicit SUSY
example of this idea.

\medskip

As should be clear, one can apply similar ideas to generate other 
interesting couplings forbidden in perturbation theory by some 
massive $U(1)$ symmetry. For instance, the Yukawa couplings $10\cdot 
10\cdot 5$ in standard $SU(5)$ GUTs, which violates the $U(1)$ symmetry of 
$U(5)$ when realized in D-brane models. Or similarly, the quark Yukawa 
couplings in D-brane constructions where right-handed quarks are realized 
as antisymmetric representations of $SU(3)$. Clearly, 
non-perturbative effects open up new possibilities for improving model 
building prospects of these constructions. 

\section{Discussion}
\label{discussion}

We have found that neutrino Majorana masses are generated by D-brane
instantons in certain general classes of string compactifications.
We have discussed a set of necessary conditions for this mechanism to be 
allowed, like the presence of a massive (although obviously 
anomaly-free) $U(1)_{B-L}$ generator. This requirement is non-trivial and 
in fact it is not satisfied in most semirealistic constructions to date. 
Hence those models should be regarded as not fully realistic in the 
neutrino sector. The Majorana mass term constraint thus turns out to be a 
powerful new ingredient in model building requirements.

Although non-automatic, the requirements are however satisfied by a 
restricted but non-trivial subset of models. To our knowledge,
only the examples \cite{imr} discussed above and some SUSY models
built by Schellekens and collaborators \cite{schell} using CFT techniques 
in
Type II orientifolds have this property. It would be very interesting and  
important to construct new  models with this property using
different string constructions. In the heterotic case that  will require 
using $U(N)$ gauge bundles for compactification rather than $SU(N)$. 

One point to remark is that, even within that class of models,
the existence of the required instanton with the appropriate number of 
fermionic zero modes is also a strong constraint. For instance, in our 
toroidal orientifold example among the large class of three generation 
models which one can build with the wrapping numbers of Table 2, only 
those satisfying the constraints $|n_c^1|=1, n_a^2=$ multiple of 3 
have the appropriate zero modes to obtain Majorana neutrino masses. 
As we saw, this may have  a bearing on the possible number of generations 
and of Higgs multiplets in this class of models. More generally, we
expect that imposing the existence of the required D2 instanton in generic 
constructions may give constraints  on the number of 
generations and/or Higgs multiplets. It should be interesting
to check for those constraints in general models.

The finding of this new source for the generation of 
Majorana neutrino masses opens the way to the study 
of the neutrino sector in string models. As in the case of the
masses and mixings of quarks and charged leptons, they will
depend on the details of the string compactification
(brane geometry in the case of an intersecting brane setup).
On the other hand the conditions that we have found for the generation 
of neutrino masses are topological in nature and hence are
much easier to implement in a systematic search for 
a string vacuum consistent with phenomenological data.
For example, one may consider the class of CFT MSSM-like models constructed 
in \cite{schell}. To begin with one should then concentrate on the
limited set of models with a massive $U(1)_{B-L}$ and then look on
whether one may find branes (corresponding to the instantons) 
with the appropriate intersection numbers (\ref{condi}) with the SM 
branes. That would single out a direction to go in the search for a fully 
realistic MSSM-like model.

As we argued, an additional important aspect of the proposed
Majorana mass generation mechanism is that it is such that a
$Z_2$ subgroup of the massive $U(1)_{B-L}$ generator remains
unbroken. In the context of the MSSM this is equivalent to the
existence of R-parity, which guarantees the absence of dimension four
operators violating baryon or lepton number, a crucial ingredient of the MSSM
which is imposed by hand in field theory. We now see that string theory
would provide  a rationale  for the existence of this symmetry
which is connected to the generation of neutrino masses.

String instantons may also give rise to other interesting 
superpotential terms in the low effective action. An
important example that we have discussed in the text is the Higgs bilinear,
the $\mu $-term in the MSSM. Dimension four operators may also be 
obtained. For example, it is often the case  in specific  semirealistic
D-brane models that some potential Yukawa coupling is forbidden by 
some anomalous $U(1)$ symmetry (like e.g. $U(1)_b$). Instanton effects
may generate such couplings although the size of those terms 
will be generically exponentially suppressed compared to other
allowed Yukawa couplings. This may be perhaps interesting in connection
with the generation of hierarchies of quark/lepton masses.

Coming back to neutrino masses, it is clear 
that obtaining  some specific  prediction for the masses and mixings of
neutrinos requires the study of concrete models. However  one can argue that
for a string instanton generation of neutrino Majorana masses 
it is natural to expect large neutrino mixing, 
 having small mixings 
would be rather surprising. Indeed, at least from the intuition provided 
by string intersecting brane  models, one may 
perhaps understand qualitatively why the mixing among quark flavours 
given by the CKM matrix is relatively small. That may happen e.g.if there
is some approximate left-right symmetry in the geometric distribution of 
D6-branes.
 The CKM matrix is related to the unitary matrices which diagonalize 
the quark masses and the Yukawa couplings depend on the geometry of the
wrapping SM branes. The neutrino Dirac mass matrix $M_\nu^D$ will 
also depend on the geometry of the SM D-branes. However 
we have seen that 
the origin of the right-handed Majorana neutrino mass
is totally different, not only depends on the geometry of some SM branes
but also on that of the instanton M generating the coupling. Since the physically
 measured  light neutrino masses depend on the Dirac neutrino mass matrix as well
as on the right-handed Majorana mass, having both matrices totally distinct
origin in our scheme, no particular correlation is expected which
generically implies large neutrino mixing, as experimentally observed.

\vspace*{1cm}

{\bf \large Acknowledgments}

We thank A. Font and  F. Marchesano  for useful discussions.
L.E.I. thanks CERN's PH-TH Division for hospitality while
this work was being carried out. A.M.U. thanks M. Gonz\'alez for kind 
encouragement and support. 
This work has been partially supported by the European Commission under
the RTN European Programs MRTN-CT-2004-503369, MRTN-CT-2004-005105, 
the CICYT (Spain), and the Comunidad de Madrid under project HEPHACOS, 
P-ESP-00346.

\newpage

\appendix

\section{String theory instantons and effective operators}
\label{generalinst}

\subsection{D-brane models}

There are many discussions of brane instanton physics in string and M-theory
in the literature (see e.g. 
\cite{Becker:1995kb,Witten:1996bn,Harvey:1999as, Witten:1999eg} among others). However 
they usually do not many deal with models with non-trivial gauge sectors 
and matter charged under them. Hence, the appearance of instanton induced 
operators of the kind we are interested in has not been much discussed, 
In this appendix we extend results in \cite{Ganor:1996pe,mcgreevy} 
and discuss the microscopic mechanism for euclidean 
D-brane instantons to generate effective operators involving the charged 
fields in a string compactification. The language is completely general, 
and it is straightforward to particularize to any compactification (either 
in type IIA or type IIB) with D-branes. For simplicity we ignore 
orientifold projections, which can be easily incorporated (as done in the 
discussion in the main text). Despite of this and as discussed in the main 
text, we use supersymmetric language; in any event the relevant instanton 
physics is independent of supersymmetry. Also, as discussed in section 
\ref{extra}, we focus on the relevant instanton zero modes, 
ignoring the possible presence of additional ones.

Consider a compactification of type IIA or IIB string theory with D-branes 
leading to four-dimensional gauge interactions and charged chiral 
fermions. This could be a type IIA compactification with D6-branes on 
intersecting 3-cycles \cite{bgkl,afiru,interev}, a type IIB 
compactification with magnetized D-branes \cite{bachas,bgkl,aads}, a type 
IIB model with D-branes at singularities \cite{aiqu,alday,delta}, or 
even non-geometric compactifications like orientifolds of type II Gepner 
models \cite{gepner2,gepner1,schell}. 
In fact, since the ingredients are essentially topological, diverse 
dualities can be used to draw similar conclusion in other setups, like 
M-theory on $G_2$ holonomy manifolds, or F-theory on CY fourfold.
The case of heterotic models is particularly interesting and will be 
discussed in section \ref{heterotic}.

As is by now familiar, we have several stacks, labeled by an index $A$, of 
$N_A$ D-branes, denoted $D_A$ branes. Each D-brane is characterized by a 
vector of RR charges $\Pi_A$. This corresponds to the homology charge of 
the D-brane in geometric compactifications, or to a suitable 
generalization in other models. These charge vectors admit a decomposition 
in a basis of D-brane charges $C_r$ as follows
\beqa
\Pi_A \, =\, \sum_r\, p_{Ar} \, C_r
\eeqa
where $p_{Ar}$ are integers. The basis of D-brane charges $C_r$ is 
associated to a basis of RR forms in the 4d theory, which in geometric 
compactification corresponds to cohomology basis of the internal space 
(and to suitable generalizations in more general models). By abuse of 
language we use $C_r$ to denote the basic D-brane charge and the 
corresponding cohomology class (or suitable generalization thereof).

The $p_{Ar}$ correspond to the charge of the D$_A$-brane under the 
$r^{th}$ RR 4-form of the four-dimensional effective theory. For 
instance, in geometric compactifications, such 4-form is given by KK 
reduction of a suitable RR form of the 10d theory over the basis cycle 
associated to $C_r$. The equations of motion for these 4d 4-forms 
imply the RR tadpole cancellation constraints (recall we do not include 
orientifold planes in the discussion).
\beqa
\sum_A\, N_A\, \Pi_A \, =\, 0
\label{rrtadpole}
\eeqa
The 4d theory contains a gauge symmetry $\prod_A U(N_A)$. 
The $AB$ open string sectors provide chiral fermions (chiral multiplets in 
susy cases), with a multiplicity $I_{AB}$, in bi-fundamental representations 
$(\fund_A,\antifund_B)$, hence with charges $+1$, $-1$ under $U(1)_A$, 
$U(1)_B$. As usual, opposite signs of $I_{AB}$ indicate 
conjugate representations. The multiplicity is determined by a topological 
bilinear in the charge vectors
\beqa
I_{AB}\, =\, \langle \Pi_A, \Pi_B \rangle
\eeqa
For type IIA intersecting brane models, this corresponds to the 
topological intersection number of the 3-cycle homology classes. For 
geometric type IIB models with magnetized branes, this is the index of the 
Dirac operator for a fermions coupled to the $U(1)_A\times U(1)_B$ bundle 
(or a suitable generalization to sheaves for lower dimensional branes). 
For D-branes at singularities, it is the adjacency matrix of the quiver 
diagram (which in the large volume limit corresponds to the just mentioned 
Dirac index). For abstract CFTs, it is the bilinear described in 
\cite{Brunner:1999jq}.

\medskip

The $U(1)_A$ gauge bosons have non-trivial couplings to a set of basic 
2-forms $B_r$ (associated to the classes $C_r$) in the 4d theory, given by
\beqa
S_{BF}\, =\, \sum_{A,r} \,p_{Ar} \int_{4d}\, B_r \wedge \tr F_A\, =\, 
\sum_{A,r}\, N_A \,p_{Ar} \int_{4d}\, B_r \wedge F_A\,
\eeqa
where the factor of $N_A$ arises from the $U(1)$ generator 
normalization. This implies that, upon $U(1)_B$ gauge transformations
\beqa
A_B \to A_B \, + \, d\Lambda_B
\eeqa
the scalar $a_r$, which is the 4d dual to $B_r$, suffers a shift
\beqa
a_r \to a_r \, +\, \sum_B\,  N_B\, p_{Br}\, \Lambda_B 
\label{scalarshift}
\eeqa
In geometric compactifications, the scalar $a_r$ is obtained by 
integration over the cycle $D_r$ dual to $C_r$ of the RR form dual to that 
leading to $B_r$ (and suitable generalizations for other non-geometric 
D-brane models).

Some of the $U(1)$'s with BF couplings are anomalous and this coupling 
is crucial in the Green-Schwarz cancellation of anomalies. However, as 
emphasized in the main text, we are particularly interested in 
non-anomalous $U(1)$'s which nevertheless also have these BF couplings.

Let us now consider an euclidean D-brane instanton, namely a D-brane, 
denoted $M$, which 
is localized in all 4d Minkowski dimensions, and has a RR charge vector 
$\Pi_{M}$. Just as for 4d spacetime filling D-branes, this corresponds to 
an euclidean D2-brane on a 3-cycle on type IIA geometric models, to a 
(possibly magnetized) D$(2p+1)$-brane on a $(2p+1)$-cycle in geometric 
type IIB models, or to suitable generalizations in other setups.
Expanding $\Pi_M$ on the dual basis
\beqa
\Pi_M\, =\, \sum_r \, q_{M,r} \, D_r
\eeqa
the usual amplitude of the instanton is
\beqa
e^{-S_{inst}}\, =\, \exp(\, - (V_{\Pi_M} +i \sum_r q_{M,r} a_r)\, )
\label{instexp}
\eeqa
This would seem puzzling, since this amplitude is not invariant under 
$U(1)_a$ gauge transformations
\beqa
e^{-S_{inst}}\, \to\, \exp(\,-i\sum_{A,r} N_A q_{M,r} p_{A,r} \Lambda_A )
\, e^{-S_{inst}}\, =\, \exp(\, -i \sum_A N_A I_{M,A} \Lambda_A} \,)
\,e^{-S_{inst}
\label{expphase}
\eeqa
The puzzle is solved by the fact that the instanton in general has 
fermionic zero modes over which one should integrate. Namely, the sector 
of open strings stretching between the euclidean D-brane and the 
$A^{th}$ background D-brane leads to $I_{M,A}$ fermionic zero modes for 
the instanton (here $I_{M,a}=\langle \Pi_M,\Pi_a\rangle$), transforming in 
the bi-fundamental $(\fund_M,\antifund_A)$. It is convenient to split the 
set of $D_A$-branes into two subsets, labeled by indices $P$, $Q$, with 
$I_{M,P}$ and $I_{M,Q}$ positive and negative respectively. We also denote 
$\alpha^P_{i_P}$, $\beta^Q_{j_Q}$ the corresponding instanton fermion 
zero modes, with the indices $i_P=1,\ldots, I_{M,P}$, $j_Q=1,\ldots, 
|I_{M,Q}|$ labeling the multiplitity in a given sector. Notice that due 
to the RR tadpole cancellation (\ref{rrtadpole}), the numbers
$N_\alpha$, $N_\beta$ of positive and negative chirality fermion zero 
modes are equal 
\beqa
N_\alpha - N_\beta\, =\, \sum_{P} N_P I_{M,P} - \sum_{Q} N_Q I_{M,Q}\, 
=\, \sum_A N_A I_{M,A}\, =\, \langle \Pi_M, \sum_A N_A \Pi_A \rangle\, 
=\, 0  \quad
\eeqa

In general, the instanton zero modes have non-trivial cubic couplings 
with fields $\Phi^{PQ}_{k_{PQ}}$, with $k_{PQ}=1,\ldots, I_{PQ}$ in the 
$PQ$ open string sector of the background D-branes, of the form
\beqa
S_{z.m.}\, =\, d_{i_Pj_Qk_{PQ}} \alpha^{P}_{i_P} \beta^{Q}_{j_Q} 
\Phi^{PQ}_{k_{PQ}}
\eeqa
Integration over the Grassman variables $\alpha$, $\beta$ in the 
instanton path integral, leads to a term proportional to the determinant 
of the $N\times N$ matrix $\Phi$, with 
$N=\sum_P N_P I_{M,P}=\sum_Q N_Q I_{M,Q}$. This term, which we denote 
$(\det \Phi)$ for short, is a prefactor that accompanies the exponential 
(\ref{instexp}) in the complete instanton amplitude. Since it is 
roughly an order $N$ polynomial in fields in the $AB$ sector, under the 
$U(1)$ gauge transformations, it transforms as
\beqa
\det \Phi \to \exp(\, i \sum_P N_P I_{M,P} - i \sum_Q N_Q I_{M,Q}\, )\, 
\det\Phi\, =\, \exp (\, i \sum_A N_A I_{M,A} \Lambda_A) \, \det\Phi
\nonumber
\eeqa
which precisely cancels the transformation of the exponential, leading to 
a gauge invariant 4d interaction.

\subsection{Heterotic models}
\label{heterotic}

One can carry out a similar discussion for heterotic models (see 
\cite{hetinst} for early discussions). In fact, 
compactifications of the heterotic strings with $U(N)$ gauge bundles (as 
opposed to $SU(N)$ bundles) lead to 4d theories with a
structure of gauge factors (and most notably of $U(1)$ factors) similar 
to that in D-brane models in the previous section. This has been 
discussed in \cite{Blumenhet}. This is 
nicely consistent with S-duality of type I and the $SO(32)$ heterotic 
models. In the following we focus on instanton effects on such geometric 
$SO(32)$ heterotic constructions \footnote{One can also consider the 
$E_8\times E_8$ theory, which is conceptually similar, with differences 
only at the group-theoretical level}.

Focusing on abelian bundles, the backgrounds are most simply described by 
regarding each of the Cartan generators $Q_A$ of 
$SO(32)$ as an antisymmetric $2\times 2$ block, which plays a role 
similar to a D9-brane and its orientifold image in a type I 
compactification. Hence a non-trivial abelian field strength 2-form 
\beqa
F\, =\, \sum_{A=1}^{16} F_A
\eeqa 
(where $F$ represents an abelian $SO(32)$ matrix and $F_a$ is a $SO(32)$ 
matrix with entries only in the $a^{th}$ $SO(2)$ block) is completely 
similar to turning on a field strength $F_a$ in the $a^{th}$ D9-brane 
(and $-F_a$ in its image) in a type I model. 

Moreover, the structure of 2-forms and their dual scalars in the 
4d-theory is also similar to that of type I models. 
In the KK reduction of the heterotic string, the 10d 2-form leads 
to a universal 4d 2-form $B_0$. In addition, one can integrate the
10d 6-form over the $h_{1,1}$ independent 4-cycles $C_r$ in the 
Calabi-Yau to obtain further 4d 2-forms
\beqa
B_r\, =\, \int_{C_r}\, B_6
\eeqa
These 2-forms couple to the 4d $U(1)$ gauge fields. These couplings arise 
from the 10d Chern-Simons couplings 
\beqa
S_{CS1} & = & \int_{10d} \, B_2 \wedge \tr (F\wedge F\wedge F\wedge F)
\nonumber \\
S_{CS2} & = & \int_{10d} \, B_6\wedge \tr(F\wedge F)
\eeqa
which are crucial for the 10d Green-Schwarz mechanism. Namely, the first 
leads upon KK reduction to
\beqa
S_{CS1,4d} \, =\, N_A p_{A0}\, \int_{4d}\, B_0 \wedge F_A
\eeqa
with $p_{A0}=\int_{CY} \tr F_A^{\, 3}$. On the other hand, from the second 
kind of 10d coupling we obtain the 4d couplings
\beqa
S_{CS2,4d} \, =\, N_A\, p_{Ai}\, \int_{4d} \,  B_r \wedge F_A
\label{cs24d}
\eeqa
where 
\beqa
\int_{D_r} F_A\, =\, p_{Ar}
\eeqa
and $D_r$ is the 2-cycle dual to $C_r$.
As usual the $N_a$ factor arises from the $U(1)$ normalization.

The dual scalars $a_r$, therefore suffer a shift under $U(1)_A$ gauge 
transformations, given by (\ref{scalarshift}). For the scalars dual to the 
2-forms $B_r$, $r\neq 0$, the coupling to the $U(1)$ factors can be 
recovered in a language more familiar in the heterotic literature, as 
follows. The field strength for the 10d 2-form roughly has the structure
\beqa
H_{MNP}\, =\, \partial_{[M} B_{NP]}\, +\, A_{[M}\, F_{NP]}
\eeqa
with the additional piece required to yield the anomalous Bianchi 
identity. The mixed terms in the 10d kinetic term for $H_{MNP}$
lead to 10d couplings
\beqa
\int_{10d}\, d^{10}x\, \partial_{[M} B_{NP]}\, A_{[M}\, F_{NP]}
\eeqa
which upon KK reduction lead to the 4d couplings
\beqa
N_B \, p_{Br}\, \int_{4d} \, d^4x\, \partial_\mu a_r A_{B,\mu}
\eeqa
where $a_r=\int_{D_r} B_2$ is the scalar dual to $B_r$ above. These 
couplings are equivalent to (\ref{cs24d}), and imply the mentioned scalar 
shift.

The shift of the scalars under $U(1)$ gauge transformations renders the 
exponential amplitudes of certain instantons naively non-gauge-invariant. 
Specifically, the generic such instanton will be a bound state of an 
euclidean NS5-brane wrapped on the whole Calabi-Yau and euclidean 
fundamental strings wrapped on 2-cycles. This bound state admits an explicit 
realization as a magnetized NS5-branes, namely NS5-branes with a 
non-trivial background for its worldvolume symplectic gauge field. The 
discussion is however insensitive to this detailed realization, and only 
depends on the vector of charges $(q_0;q_i)$ of the bound state (where 
$q_0$ denotes the NS5-brane charge and $q_i$ the charge of fundamental 
strings wrapped on $D_i$. The naive amplitude of such instanton clearly 
shifts as
\beqa
e^{-S_{inst}}\, \to\, \exp(\,-i\sum_{A,r} N_A q_{M,r} p_{a,r} \Lambda_A )
\, e^{-S_{inst}}\, 
\eeqa
which is in fact identical to (\ref{expphase}).

This phase will in fact be canceled by the appearance of spacetime 
charged fields, due to integration over zero modes of the instantons. 
The microscopic description of instantons with $q_0\neq 0$ is not 
available, since it involves heterotic NS5-branes. The results for this 
can nevertheless by derived by simply dualizing results from type I 
models, which we leave as an exercise. We rather focus on the case 
$q_0=0$ which is in fact very interesting since it corresponds to a 
world-sheet instanton on the curve $D=\sum_r q_r D_r$, for which one has 
a microscopic description. Thus we can directly compute the  
field-dependent prefactor and verify the gauge invariance of the complete 
instanton amplitude.

In the fermionic world-sheet formulation of the $SO(32)$ 
heterotic, there are 32 2d fermions in the fundamental of $SO(32)$, which 
couple to the spacetime gauge field $A$ in the adjoint as 
$A\lambda\lambda$. Let us label as $\lambda^B$, $\lambda^{B^*}$, 
$a=1,\ldots 16$, and split the $SO(32)$ adjoint field accordingly. Then 
the coupling becomes
\beqa
A^{BC} \lambda^B \lambda^C \, +
A^{B^*C} \lambda^{B^*} \lambda^C \, +
A^{BC^*} \lambda^B \lambda^{C^*} \, +
A^{B^*C^*} \lambda^{B^*} \lambda^{C^*}
\eeqa
The analogy with an euclidean D1-brane instanton in a type I model should 
be clear at this point. In fact, the KK reduction of the $SO(32)$ gauge 
field in a given sector e.g. $AB$, lead to $4d$ chiral fields 
$\phi^{AB}_{k_AB}$ in the corresponding $AB$ sector. Here the label 
$k_{AB}=1,\ldots, I_{AB}$ takes into account the multiplicity $I_{AB}$ 
given by the index of the Dirac operator for a field with charges $\pm 
1$ under $U(1)_A\times U(1)_B$. Also, the KK reduction of the 2d 
fermions e.g. $\lambda_A$ lead to $I_{M,A}$ instanton fermionic zero 
modes, where $I_{M,A}$ is the index 
of the Dirac operator for a field with charges $+1$ under $U(1)_A$. Hence 
the instanton contains cubic couplings among the $I_{MA}$, $I_{MB}$ 
fermion zero 
modes $\alpha$, $\beta$ and the spacetime fields $\phi$. The integration 
over fermion zero modes leaves a determinant in the latter, whose 
transformation cancels the phase of the exponential, as should be 
familiar by now.

An interesting point to emphasize is that, for $q_0=0$, the relevant 
instantons are world-sheet instantons, hence they are not suppressed by 
$g_s$. Rather they are tree-level in the string coupling, but 
non-perturbative in $\alpha'$.

\section{Some further intersecting brane examples}
\label{moremodels}

In this appendix we describe how neutrino Majorana mass terms may appear 
in the family of models considered in \cite{imr}. We refer to that paper 
for notation and details. One important difference with the family of 
models in the main text is that  the $SU(2)_L$ gauge group comes from a 
$U(2)_b$ group and the number of generations is fixed to three from the 
start. The wrapping numbers of the SM D6-branes in this family of models 
are given by Table \ref{solution}.
%
\begin{table}[htb] \footnotesize
\renewcommand{\arraystretch}{2.5}
\begin{center}
\begin{tabular}{|c||c|c|c|}
\hline
 $N_i$    &  $(n_i^1,m_i^1)$  &  $(n_i^2,m_i^2)$   & $(n_i^3,m_i^3)$ \\
\hline\hline $N_a=3$ & $(1/\beta ^1,0)$  &  $(n_a^2,\epsilon \beta^2)$ &
 $(1/\rho ,  1/2)$  \\
\hline $N_b=2$ &   $(n_b^1,-\epsilon \beta^1)$    &  $ (1/\beta^2,0)$  &
$(1,3\rho /2)$   \\
\hline $N_c=1$ & $(n_c^1,3\rho \epsilon \beta^1)$  &
 $(1/\beta^2,0)$  & $(0,1)$  \\
\hline $N_d=1$ &   $(1/\beta^1,0)$    &  $(n_d^2,-\beta^2\epsilon/\rho )$  &
$(1, 3\rho /2)$   \\
\hline \end{tabular}
\end{center} \caption{ D6-brane wrapping numbers giving rise to a SM spectrum
as in ref.\cite{imr}.
\label{solution} }
\end{table}
%
The models are parametrized by a phase $\epsilon =\pm1$, four integers
$n_a^2,n_b^1,n_c^1,n_d^2$ and a parameter $\rho=1,1/3$. In addition 
$\beta^i=1-b^i=1,1/2$ depending on whether the corresponding tori are
tilted or not. Such classes of models have in principle up to four $U(1)$ 
gauge fields, but generically three of them acquire St\"uckelberg masses  
due to the $B\wedge F$ couplings. In particular one has
\beqa
B_2^1 &\wedge & \ {{-4\epsilon \beta^1}\over { \beta^2 } }F^{b} \nonumber
\\
B_2^2 &\wedge  & \ \frac{(2\epsilon  \beta^2 )}
{\rho \beta^1}(3F^a\ -\  F^{d})
\nonumber \\
 B_2^3 &\wedge  &  \ {1\over { \beta^2 } }
(\, \frac{3\beta^2 n_a^2}{\beta^1} F^a\ +\ 6\rho n_b^1F^{b} \ +\
2n_c^1F^c \ +\ \frac{3\rho  \beta^2 n_d^2}{\beta^1} F^d\, ) \
\label{bfs}
\eeqa
The four scalar fields $a_i$  transform with a shift
under $U(1)$ transformations as 
\beqa
a_0\ &\rightarrow \ a_0 \nonumber\\
a_1\ &\rightarrow& \ a_1\  - \ \frac {4\epsilon
\beta^1}{\beta^2}\Lambda(x)_b \nonumber \\
a_2\ &\rightarrow& \ a_2\  + \ \frac {2\epsilon \beta^2}{\rho \beta^1}
(3\Lambda(x)_{a}\ -\ \Lambda(x)_d)\\
a_3\ &\rightarrow& \ a_3\  + \frac {3n_a^2}{\beta^1} \Lambda_a(x)
\ +\
\frac {6\rho n_b^1}{\beta^2}\Lambda_b(x)
\ +\
\frac {2n_c^1}{\beta^2} \Lambda_c(x)
\ +\
\frac {6\rho n_d^2}{2\beta^1} \Lambda_d(x) \nonumber
\eeqa
There is just  one linear combination of $U(1)$'s which remains massless.
That linear combination is precisely the standard hypercharge
$U(1)_Y$ as long as one has the constraint
\beq
n_c^1 \ =\ \frac {\beta^2}{2\beta^1}(n_a^2+3\rho n_d^2)
\label{condhiper}
\eeq
It is easy to check that there is a unique factorizable 3-cycle
which may be wrapped by a D2-instanton with the required zero modes:
\beq
(n_1,m_1),(n_2,m_2),(n_3,m_3)\ =\ (3\rho n_b^1,-\beta^1),
(\rho n_a^2,\beta^2),(-\epsilon,1/2)
\eeq
as long as 
\beq
\beta^1 n_c^1 \ =\ 1 \ \ , \ \ \rho n_a^2\ =\ {\rm integer}
\eeq
Such an instanton would generate a coupling
\beq
e^{-V_{\Pi_M}} \, e^{\frac {1}{2}( 6\epsilon \rho^2n_b^1n_a^2a_0
-3\rho n_b^1 \beta^2a_1 +
\rho n_a^2\beta^1 a_2-2\epsilon \beta^1\beta^2 a_3)}
 \nu_R^i\nu_R^j \ \ ; \ \ i,j=1,2,3
\label{finaloperator}
\eeq
which is fully gauge invariant. As an example consider the 
choice of parameters

\beq
n_a^2=3\ ; n_c^1=n_d^2=\beta^1=1\ ; n_b^1=-1\ ; \rho=1/3\ ;\beta^2=1/2
\ ; \epsilon=1
\eeq
Then the relevant instanton M and branes $c^*$ and $d$ have
wrapping numbers:
\beqa
M &:& (1,1)(1,1/2)(1,-1/2)\\
c^*&:& (1,-1)(2,0)(0,-1)\\
d &:& (1,0)(1,-3/2)(1,1/2) 
\eeqa
and $I_{Md}=-I_{Mc*}=2$, $I_{Md*}=I_{Mc}=0$, as required.

\section{A $\mu $-term example}
\label{muexample}

Here we will provide an explicit MSSM-like example in which the
conditions  (\ref{condmu}) for the generation of a Higgs bilinear
are met. We will consider the   MSSM-like model
in section 6.1 of \cite{fim}, we refer the reader to that
paper and the references therein for more details. This is
a toroidal $Z_2\times Z_2$  orientifold with D6-branes with wrapping 
numbers as in table
(\ref{tablamu}).
%
\begin{table}[htb] \footnotesize
\renewcommand{\arraystretch}{2.5}
\begin{center}
\begin{tabular}{|c||c|c|c|}
\hline
 $N_i$    &  $(n_i^1,m_i^1)$  &  $(n_i^2,m_i^2)$   & $(n_i^3,m_i^3)$ \\
\hline\hline $N_a=6+2$ & $(1,0)$  &  $(3, 1)$ &
 $(3,-1/2)$  \\
\hline $N_b=4$ &   $(1,1)$    &  $ (1,0)$  &
$(1,-1/2)$   \\
\hline $N_c=2$ & $(0,1)$  &
 $(0,-1)$  & $(2,0)$  \\
\hline
\hline $N_X=4$ &   $(-2,1)$    &  $ (-3,1)$  &
$(-3,1/2)$   \\
\hline $N_O=6$ &   $(1,0)$    &  $ (1,0)$  &
$(1,0)$   \\
\hline \end{tabular}
\end{center} \caption{ D6-brane wrapping numbers giving rise to the 
MSSM-like
model in the text.
\label{tablamu} }
\end{table}
In order to cancel RR-tadpoles one can add a single coisotropic D8-brane 
stack
as in \cite{fim} or else two D6-brane stacks as in table 
\ref{tablamu},
it is not relevant for the present discussion.
The gauge group after one takes into account those $U(1)$'s getting 
St\"uckelberg masses
is  that of the SM plus an  additional $B-L$ (and a $SU(2)_X$ 'hidden 
sector'
group). Since in this model $B-L$ is massless
the generation of a Majorana neutrino mass operator is not possible.
The spectrum may be found in table 5
of \cite{fim} (the multiplicity of the $G,{\bar G}$ states in that table 
is 6 instead
of 5 if we use D6-branes X,O instead of a D8-brane). It corresponds to a 
three generation
MSSM-like spectrum with a minimal set of Higgs fields $H$,${\bar H}$ plus 
additional
chiral fields transforming under the electroweak and the $SU(2)_X$ 'hidden 
group'.
It can be checked that this model preserves $N=1$ SUSY and all RR-tadpoles 
cancel.
It is easy to check that  a D2 instanton wrapping the 3-cycle M
\beq
M\ =\ 2(1,0)(1,-1)(1,1/2)
\eeq
has the appropriate intersection numbers in eq.(\ref{condmu}),
i.e. $I_{Mb} = -1$,$I_{Mb*} = 0$,$I_{Mc}=I_{Mc}* = 1$. In addition it
preserves the same $N=1$ SUSY as the D6-branes. One has 
$\im S_{D2}=a_0-\frac{1}{2}a_1$
so that   under
a $U(1)_b$ gauge transformation of parameter $\Lambda _b(x)$ one has
\beq
S_{D2}\ \longrightarrow \ S_{D2}\ -\ i2\Lambda_b(x)
\eeq
so that indeed the operator $exp(-S_{D2})H{\bar H}$ is gauge invariant 
under
$U(1)_a,U(1)_b$, and $U(1)_c$ as expected. However one can check that in 
the
present example there are non-vanishing intersections $I_{MX} = -4$ ,
$I_{MX*} = 4$  between the instanton and the auxiliary branes $X$ (or 
their coisotropic
D8-brane analogues) which are added to cancel RR-tadpoles. This implies 
that
$exp(-S_{D2})$ is also charged under $U(1)_X$ and that the actual operator 
which
could be induced will have and additional factor involving fields charged
under the 'hidden sector' gauge group $U(2)_X$.

\end{document}